\documentclass[%
 reprint,
superscriptaddress,
 amsmath,amssymb,
prb,
]{revtex4-2}

\usepackage{graphicx}% Include figure files
\usepackage{dcolumn}% Align table columns on decimal point
\usepackage{bm}% bold math
\usepackage{appendix}
\usepackage{soul}
\usepackage[utf8]{inputenc}
\usepackage[english]{babel}
\usepackage[T1]{fontenc}
\usepackage[colorlinks,bookmarks=false,citecolor=blue,linkcolor=red,urlcolor=blue]{hyperref}
\usepackage[dvipsnames]{xcolor}
\usepackage{times}
\usepackage{bbm}
\usepackage[normalem]{ulem}
\usepackage{multirow}
\usepackage{booktabs}
\usepackage{amssymb}
\usepackage{amsmath}

\usepackage[final]{changes}

\definechangesauthor[name = JF, color=blue]{JF}
\definechangesauthor[name = GK, color=red]{GK}
\definechangesauthor[name = JP, color=orange]{JP}

\newcommand{\be}{\begin{eqnarray}}
\newcommand{\ee}{\end{eqnarray}}

\begin{document}

\title{Thermodynamic signatures of short-range magnetic correlations in UTe$_2$}

\author{Kristin~Willa}
\affiliation{Institute for Quantum Materials and Technologies, Karlsruhe Institute of Technology, 76021 Karlsruhe, Germany}

\author{Frédéric~Hardy}
\affiliation{Institute for Quantum Materials and Technologies, Karlsruhe Institute of Technology, 76021 Karlsruhe, Germany}

\author{Dai~Aoki}
\affiliation{Institute for Materials Research, Tohoku University, Ibaraki 311-1313, Japan}
\affiliation{Univ. Grenoble Alpes, CEA, Grenoble INP, IRIG, PHELIQS, F-38000 Grenoble, France}

\author{Dexin~Li}
\affiliation{Institute for Materials Research, Tohoku University, Ibaraki 311-1313, Japan}

\author{Paul~Wiecki}
\affiliation{Institute for Quantum Materials and Technologies, Karlsruhe Institute of Technology, 76021 Karlsruhe, Germany}

\author{Gérard~Lapertot}
\affiliation{Univ. Grenoble Alpes, CEA, Grenoble INP, IRIG, PHELIQS, F-38000 Grenoble, France}

\author{Christoph~Meingast}
\affiliation{Institute for Quantum Materials and Technologies, Karlsruhe Institute of Technology, 76021 Karlsruhe, Germany}

\date{\today}
 
\begin{abstract}
The normal-state out of which unconventional superconductivity in UTe$_2$ emerges is studied in detail using a variety of thermodynamic and transport probes.  Clear evidence for a broad Schottky-like anomaly with roughly R ln 2 entropy around $T^{*} \approx 12$~K is observed in all measured quantities. Comparison with high magnetic field transport data allows the construction of an $H\text{-}T$ phase diagram resembling that of the ferromagnetic superconductor URhGe. The low field electronic Gr\"uneisen parameter of $T^{*}$ and that of the metamagnetic transition at $H_m \approx 35$~T are comparable pointing to a common origin of both phenomena. Enhanced Wilson and Korringa ratios suggests that the existence of short range ferromagnetic fluctuations cannot be ruled out. 

\end{abstract}

\maketitle

The recent discovery of superconductivity~\cite{Ran19, Aoki19-2} (SC) in the heavy-fermion UTe$_2$ has led to a tremendous amount of experimental and theoretical work. At first glance,  UTe$_2$ shares several  phenomena with the well established ferromagnetic superconductors UGe$_2$, URhGe and UCoGe~\cite{Aoki19}. These phenomena include  metamagnetism,~\cite{Miyake19,Knafo19} field-induced reentrance/reinforcement of SC~\cite{Knebel19,Knebel2020,RanNatPhys}, multiphase SC~\cite{Aoki2020,Braithwaite2019} and pressure-induced magnetic phases \cite{Braithwaite2019, Thomas2020}. Considering its low critical temperature $T_{sc}$ $\approx$ 1.6~K, the persistence of superconductivity up to magnetic fields far above the Pauli limit strongly suggests that the Cooper pairs in UTe$_2$ condense into a spin-triplet $p$-wave state~\cite{Knebel2020}. Furthermore, its large susceptibility makes magnetic fluctuations a likely pairing glue. Many questions remain, however, concerning the normal electronic state out of which superconductivity emerges, as no evidence of long-range magnetic order has been observed so far. While neutron-scattering experiments revealed incommensurate antiferromagnetic (AF) fluctuations~\cite{Duan20} along a wave vector that connects quasi-2D Fermi-surface sheets~\cite{Miao20}, nuclear magnetic resonance~\cite{Tokunaga19} (NMR) and muon spin relaxation~\cite{Sundar19} ($\mu$SR) measurements reported the existence of low-frequency longitudinal fluctuations along the $a$-axis at low temperature, although the possible proximity to a ferromagnetic instability was unclear from these data. Applying a magnetic field along the initial hard $b$-axis leads to a metamagnetic transition at $H_{\rm m} \approx 35$~T \cite{Miyake19, Knafo19, RanNatPhys} which is different to many other heavy fermion compounds where metamagnetism mostly occurs for a magnetic field applied along the easy axis \cite{Taufour2010, Aoki19, Aoki2011}. At low temperatures, UTe$_2$ is a heavy Fermi liquid with an enhanced Sommerfeld coefficient $\gamma = 0.12$~Jmol$^{-1}$K$^{-2}$ and a $T^2$ temperature dependence of the resistivity $\rho$ below  $T\approx  5$~K \cite{Aoki19}.
Under pressure the Kondo coherence is suppressed \cite{Ran20} and multiple superconducting phases, as well as a high-pressure, most likely antiferromagnetic, phase have been found \cite{Braithwaite2019, Knebel2020, Aoki2020, Thomas2020}.

In this paper, the normal state of UTe$_{2}$ is studied by thermodynamic (thermal-expansion, specific-heat, magnetostriction and susceptibility) measurements. In all of these quantities, clear evidence for a broad Schottky-like anomaly around $T^{*}=12$~K is observed with a calculated entropy of roughly $R\ln 2$. Combining our thermodynamic measurements with previously published high magnetic-field transport data~\cite{Knafo19} allows us to construct an $H\text{-}T$ phase diagram which resembles that of the ferromagnetic superconductor URhGe \cite{Aoki19, Miyake09}. 
From our thermodynamic data, we calculate the electronic Gr\"uneisen parameter related to $T^{*}$. It is found to be comparable in magnitude to that of the metamagnetic field \cite{Li2021} indicating a common origin. Finally, enhanced Wilson and Korringa ratios suggest that the existence of short range ferromagnetic fluctuations cannot be ruled out, although no direct evidence for their existence has been found to date.
 
\begin{figure}[!t]
\centering
%-----------------------------------------------------------
\begin{minipage}{1\linewidth}
\centering
\includegraphics[width=\linewidth]{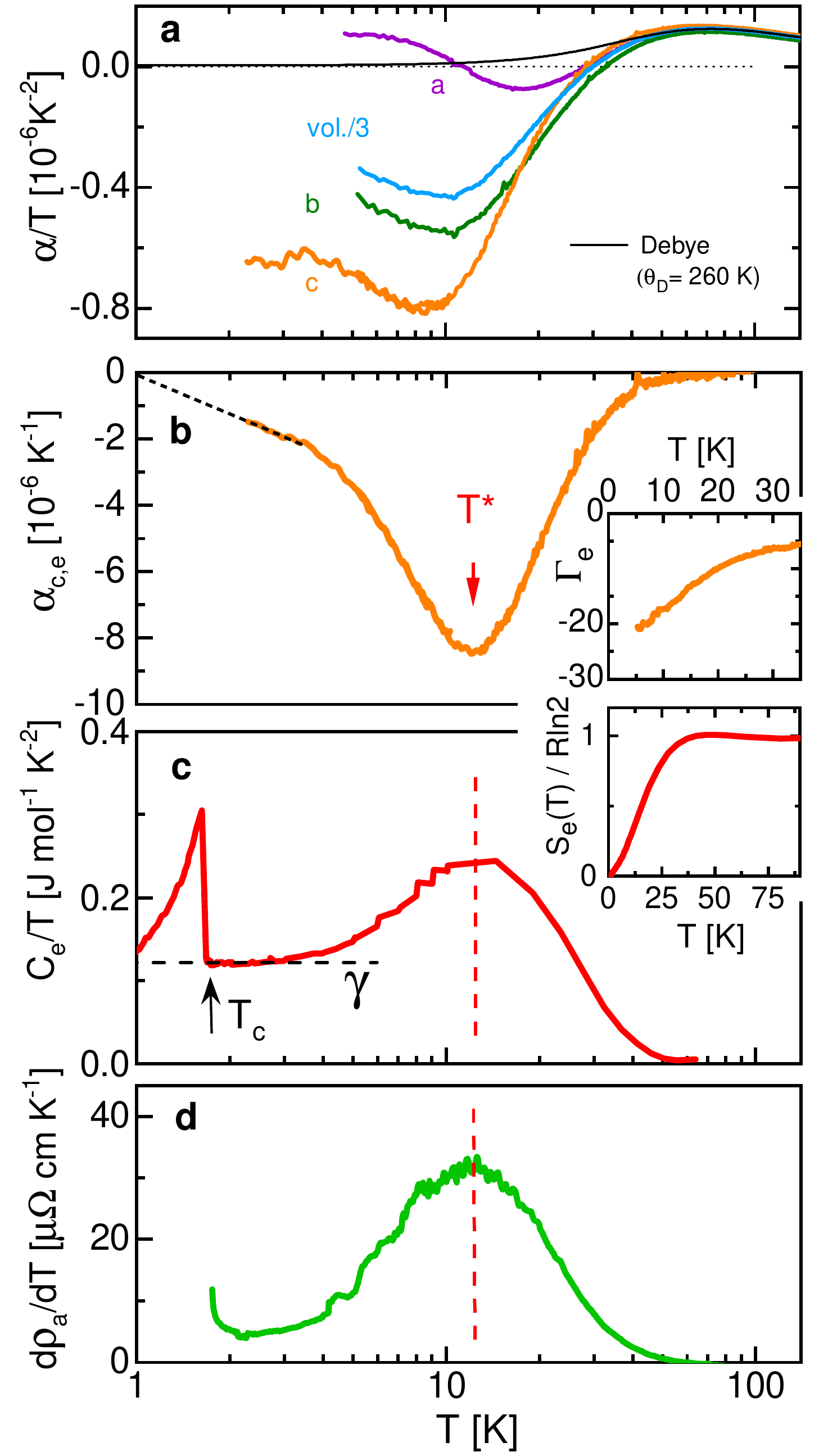}
\end{minipage}
%-----------------------------------------------------------
\caption{\textbf{a} $\alpha/T$ for all three crystallgraphic directions together with the calculated volume expansion. The solid black line is a Debye fit to the phonon background with $\Theta_D=260$~K. \textbf{b} Electronic contribution to the $c$-axis thermal expansion of UTe$_2$. The inset shows the electronic Gr\"uneisen parameter. \textbf{c} Electronic contribution of the specific heat with $\gamma$ and $T_c$ indicated with the corresponding entropy shown in the inset. Panel \textbf{d} shows the temperature derivative of the $a$-axis resistivity. The arrow and the red dashed line indicate the position of the crossover temperature $T^*$.}      
\label{Fig1}
\end{figure}

Millimeter-sized single crystals were grown by chemical vapor transport as reported in Ref.~\onlinecite{Aoki19-2}. Thermal expansion and magnetostriction were measured using  home-built dilatometers \cite{Meingast1990}.  Specific heat and magnetization were obtained with a Physical Property Measurement System (PPMS) from Quantum Design. 

The temperature dependence of the thermal expansion coefficient $\alpha_i$ along the three orthorhombic axes ($i = a, b, c$) is displayed in Figure \ref{Fig1}\textbf{a}. At high temperature $\alpha_i/T$ is small and positive and is dominated by the phonon contribution. However, it is found to change sign around 30~K for all directions, roughly where the $b$-axis susceptibility is maximum, revealing a strongly temperature- and uniaxial pressure-dependent electronic/magnetic contribution. Prominent minima are observed for all directions below 20~K, with the strongest response for $\alpha_c$, which shows a pronounced minimum at $T^* \approx 12 K$. Slight differences in the temperatures of the minima along the different axes clearly indicate a crossover phenomenon at $T^*$. Our data are in good agreement with those of Ref.~\onlinecite{Thomas2021}.

Figure~\ref{Fig1} also shows the electronic contributions to the $c$-axis thermal expansion coefficient $\alpha_{c,e}$ in panel \textbf{b}, and to the heat capacity over temperature $C_e/T$ in panel  \textbf{c} (lattice background subtraction shown in figure S1 of the Supplemental Material) as well as the derivative of the resistivity $d\rho_a/dT$ (panel \textbf{d}). The striking feature in all of these curves -- which, except for $d\rho_a/dT$, represent derivatives of the free energy -- is the broad Schottky-like anomaly centered at $T^*=12$~K. Remarkably, this anomaly also appears in $d(\chi_a-\chi_b)/dT$ and in $d\chi_a/dT$ (see Supplemental Material), which is proportional to the $H$-derivative of the electronic/magnetic entropy via a Maxwell relation. We note that recent resistivity measurements (for currents applied along the $c$ axis) and thermopower measurements also show a pronounced maximum at $T^*$ \cite{Cairns2020, Eo2021, Niu20}. With the present measurements, we are able to show that this maximum is closely related to thermodynamic bulk properties of UTe$_2$.

The large values of the electronic heat capacity of UTe$_2$ with $\gamma$ = 122 mJ mol$^{-1}$K$^{-2}$ ~\cite{Aoki19-2} and the Pauli-like susceptibility with $\chi_a$(0) $\approx$ 0.045 emu mol$^{-1}$ in 1 T (see Fig. \ref{Fig3}) provide solid evidence for the existence of a moderately heavy Fermi-liquid-like state just above the superconducting transition.  This is reinforced by the linear temperature dependence of $\alpha_c$ for T $<$ 4 K (see figure \ref{Fig1}\textbf{b}) and by thermopower data \cite{Niu20} as well as by the $T^2$ temperature dependence of the resistivity below $T \approx 5$~K \cite{Aoki19}. Interestingly, the integrated entropy of the electronic heat capacity approximately amounts to $R\ln 2$ for $T>50$ K (see inset of Fig.~\ref{Fig1}\textbf{c}).

\begin{figure*}[!t]

%-----------------------------------------------------------

\centering
\includegraphics[width=1\textwidth]{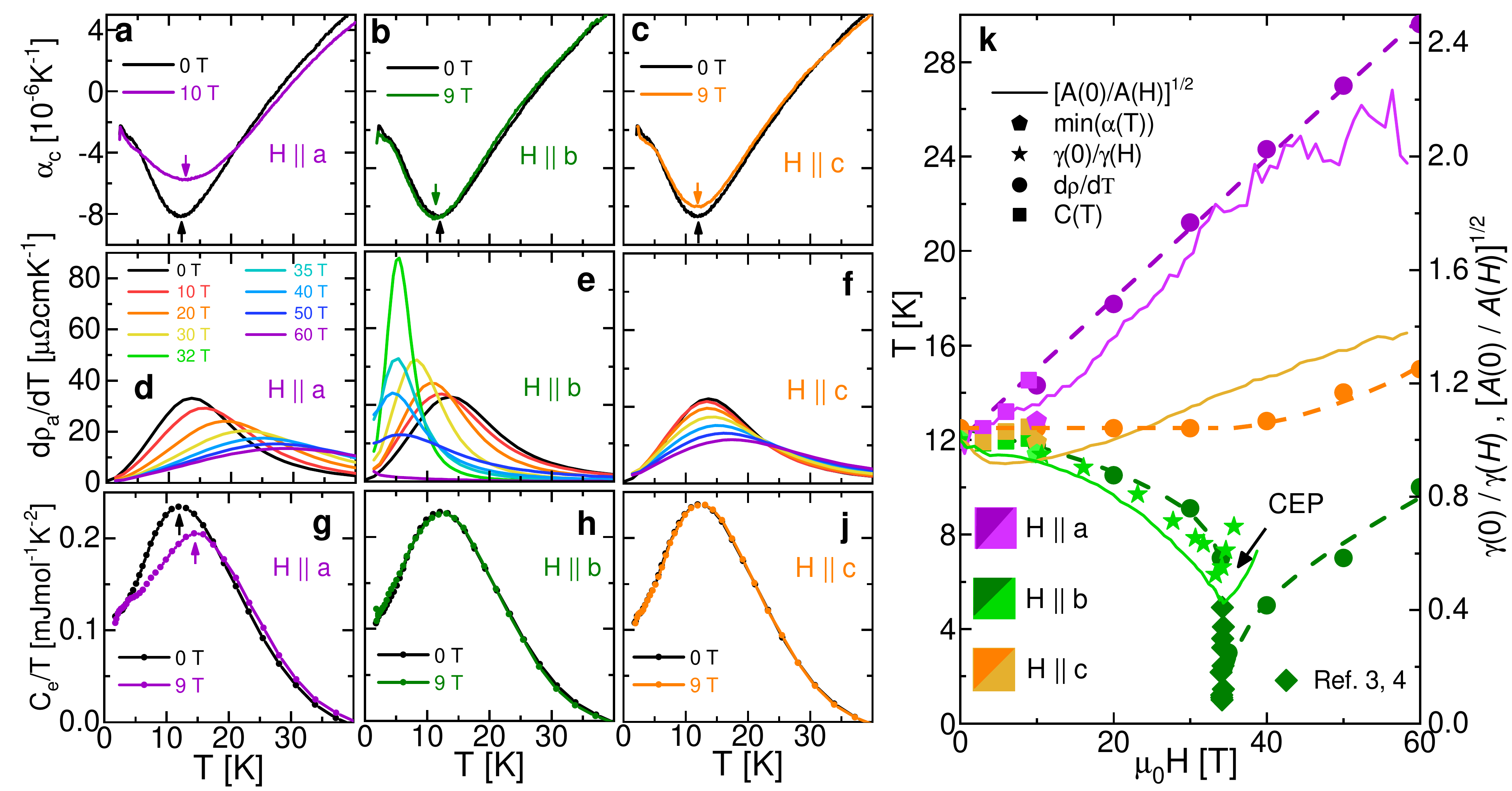}

%-----------------------------------------------------------
\caption{\textbf{a-c} temperature dependence of $\alpha_c$ in 0T and 10T applied magnetic fields \textbf{d-f} derivative of the fit to the high field resistivity data from \cite{Knafo19} (See supplementary material) \textbf{g-j} electronic specific heat in 0T and 10T. \textbf{k} shows the evolution of $T^*$(H) (circles) for the three field directions. $T^*$(H) is either determined by the position of the maximum in $d\rho_a/dT$, the maximum of $C_e$ or the minimum in $\alpha_c$  shown in Supplemental Material. The square symbols represent the first-order metamagnetic transition from Refs.~\onlinecite{Knafo19,Miyake19}. Lines correspond to the normalized field dependence of $1/\sqrt{A}$ and $1/\gamma$, taken from Ref.~\onlinecite{Knafo19} and \onlinecite{Imajo19}, respectively. Dashed lines serve as a guide to the eye.}    
\label{Fig2}

\end{figure*}

\begin{figure}[!t]
\centering
%-----------------------------------------------------------
\begin{minipage}{1.0\linewidth}
\centering
\includegraphics[width=1.0\linewidth]{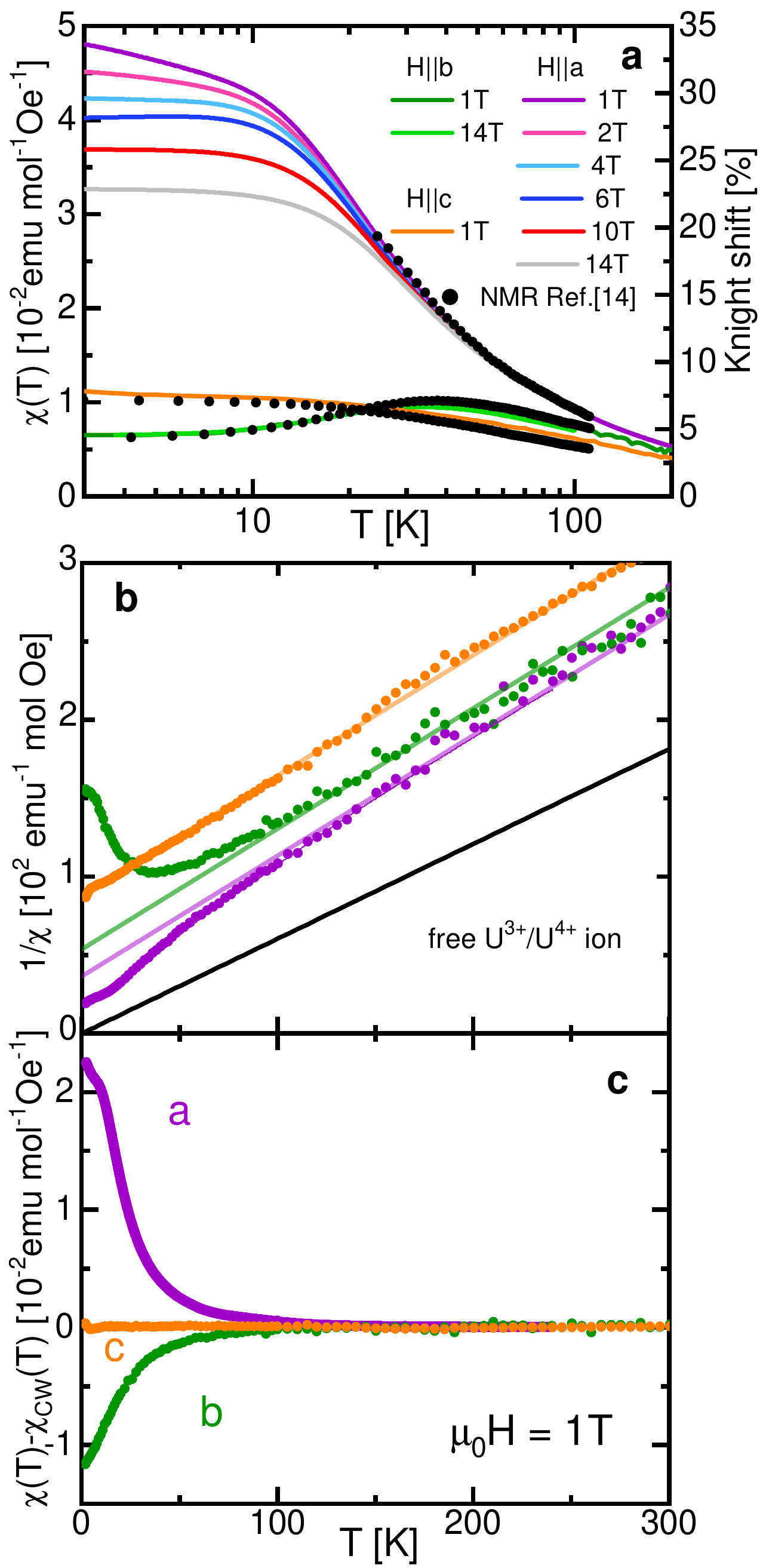}
\end{minipage}
%-----------------------------------------------------------
\caption{\textbf{a} Magnetic susceptibility measured for various fields and field directions. \textbf{b} Curie-Weiss fit for all three crystallographic directions with negative Curie-Weiss temperatures. \textbf{c} Susceptibility with subtracted Curie-Weiss fit indicating that deviations from the Curie-Weiss behavior develop together with the anisotropy between the $a$- and $b$-axis at low temperature.}
\label{Fig3}
\end{figure}

 Fig~\ref{Fig2} shows the effect of a magnetic field of 10~T applied along the different crystallographic directions upon the temperature dependence of the $c$-axis thermal expansion $\alpha_c(T)$ in Figs~\ref{Fig2}\textbf{a-c}, the temperature derivative of the resistivity as inferred from a simple fit to the published data of Knafo \textit{et al.}\cite{Knafo19} (measured at const $T$) in Figs~\ref{Fig2}\textbf{d-f} and the electronic specific heat in Figs~\ref{Fig2}\textbf{g-j}. Details about the resistivity fits are given in the Supplemental Material. The largest effect occurs along $H\parallel a$, where $T^*$ is shifted to higher temperature and the anomaly is broadened significantly similar to the behavior expected for a ferromagnetic phase transition in a field aligned with the ordered moment. For $H\parallel b$ a slight reduction of $T^*$ is observed with a slight sharpening, while only a small broadening is observed for $H\parallel c$.

As already shown in Fig. \ref{Fig1}, the position of the minimum at $T^{*}$ in $\alpha_c,e(T)$ (maximum in $C_e/T$) matches that of the maximum in $d\rho_a/dT$. 
This correspondence persists up to 10~T (9~T), the largest field for our dilatometry (specific heat) setup. This suggests that one can use the high-field resistivity data of Knafo {\it et al.}~\cite{Knafo19} to track the evolution of $T^{*}(H)$ to higher fields. Note that the derivative of $\rho$ typically resembles the specific heat and thermal expansion anomaly around a magnetic transition \cite{Fisher68, Geldart75, Meingast2009} and also around a coherence-incoherence crossover e.g. AFe$_2$As$_2$ [A={Cs, Rb, K}] \cite{Hardy2016, Hardy2013, Wiecki2020, Wiecki2021}.
The results of this analysis are plotted in the phase diagram of Fig.~\ref{Fig2}\textbf{k} together with the relative field dependence of $1/\gamma$ (from Ref.~\cite{Imajo19}) and of $1/\sqrt{A}$ (from  Ref.~\cite{Knafo19}). For $H\parallel c$, $T^{*}$ is rather insensitive to the magnetic field. In fields up to 60 T along the $a$-axis, we observe a continuation of the shift to higher temperature both in the maximum of $d\rho_a/dT$ and in $1/\sqrt{A}$. This behavior is very reminiscent of the easy-axis behavior of $T^{\scriptscriptstyle\mathrm{Curie}}$ in a ferromagnet, in which a field along the easy axis broadens and shifts the transition to higher temperatures. This effect can be also observed in the susceptibility in Fig. \ref{Fig3}.  In contrast, for $H\parallel b$, $T^{*}(H)$ shifts to lower temperature and the cross-over sharpens significantly for fields up to 32 T. On approaching the first-order metamagnetic transition  $T^*(H)$ vanishes by an almost vertical line, while both, $1/\sqrt{A}$ and $1/\gamma$, show a very pronounced minimum. The (crossover) line in the partly polarized regime above $H_m$ is very different when plotted from $1/\sqrt{A}$ or from $1/\gamma$ and might indicate that several energy scales now come into play.
This phase diagram of UTe$_2$ is surprisingly very similar in form to that of URhGe \cite{Aoki19, Miyake09}, if one replaces $T^{*}(H)$ by $T^{\scriptscriptstyle\mathrm{Curie}}(H)$.  Thus, although UTe$_2$ does not order ferromagnetically, it shares numerous key properties with URhGe, i.e. 
ferromagnetic-like behavior for fields along the easy axis, a metamagnetic transition for fields along the intermediate axis, and  inconspicuous behavior along the hard axis. Further, the reentrance/reinforcement of superconductivity at a metamagnetic transition~\cite{Knebel19}, related to the field-induced increase of $\gamma(H)$~\cite{Imajo19} , appears very similar in both compounds. From the phase diagram it is quite natural to assume that short-range magnetic fluctuations are involved in the anomaly at $T^*$, which sharpens up and then turns into the first-order metamagnetic transition. The possible nature of these fluctuations will be discussed later on.
 We note that we do not find any sign of an anomaly around 35K in any of our thermodynamic measurements suggesting that the broad maximum observed in $\chi_b(T)$ near 35 K correlates with the high-temperature tail of the 12 K anomaly as is visible after subtracting a Curie-Weiss background as explained below.

A diverging electronic Gr\"{u}neisen parameter,~\cite{Flouquet05}
\begin{equation}\label{Eq2}
\Gamma_e=B_T\frac{\partial \ln T^*}{\partial p}=B_T\frac{\alpha_e}{C_e},
\end{equation}
can serve as a smoking-gun indicator for the proximity to a pressure-induced quantum critical point~\cite{Garst2005,Zhu2003}. The result is plotted in the inset of ~\ref{Fig1}\textbf{b} where we used  the bulk modulus $B_T = 50$ GPa from Ref~\onlinecite{Honda}. $\Gamma_e$ increases rather slowly with decreasing T and extrapolates to a value of about $-25$. The monotonic temperature dependence of $\Gamma_e$ and in particular the absence of an anomalous behavior at $T^*$ suggests a single energy scale responsible for both $T^*$ and $\gamma$.
Using equation (\ref{Eq2}) and $\Gamma_e|_{12K}=-16$ we find $\partial T^* /\partial p$ = - 4~K/GPa and hence a suppression of $T^*$ to 0 around $p = 3$~GPa in agreement with recent transport \cite{Ran20} and also magnetization measurements \cite{Li2021} where this transition is observed at $p \approx 1.5$~GPa. Our value of $\Gamma_e$ is close in amplitude and sign to $\Gamma_H=-21$ related to the metamagnetic field from Ref. \onlinecite{Li2021} confirming that the suppression of $T^*$ and $H_m$ occur in the same pressure range, suggesting a common origin of both phenomena.

Additionally we have conducted susceptibility measurements along all three crystallographic axes shown in the upper panel of Fig. \ref{Fig3}. Measurements in fields of 1T along all three crystallographic axes are consistent with already published data \cite{Miyake19}. For fields along the $a$-axis, $\chi_a$ flattens at low temperatures indicative of a Pauli like susceptibility and its suppression with increasing field is suggestive of the suppression of ferromagnetic fluctuations as emphasized here above for the resistivity. There is no change in the b-axis susceptibility between 1T and 14T. Above roughly 50 K, all susceptibilities follow a Curie-Weiss dependence with a negative intercept, see Figure \ref{Fig3}\textbf{b}, implying the presence of antiferromagnetic correlations above $T^*$. Subtracting the Curie-Weiss fit leaves us with the susceptiblity shown in figure \ref{Fig3}\textbf{c}. Deviations from this Curie-Weiss behavior as well as the building of an anisotropy between the $a$- and $b$-axis are clearly associated with $T^*$. A suppression of $T^*$ by pressure will thus likely result in antiferromagnetic ordering, as has been postulated~\cite{Thomas2020}.

A sensitive indicator for the proximity to ferromagnetism is a large Wilson ratio ($R_W \gg 1$) at low T,
\begin{equation}\label{Eq1}
R_W= \frac{\pi^2 k_B^2}{\mu_0 \mu^2}\left(\frac{\chi_p}{\gamma}\right),
\end{equation}
which we calculate with our values of $\gamma$ and $\chi_p$. The result ranges between 6 and 25 calculated using $\mu=3.6\mu_B$ (free ion moment) and $\mu=g\mu_B$ ($g=2$ the gyromagnetic ratio of the free electron) respectively and is thus similar in magnitude to the value reported for the nearly ferromagnetic metal Sr$_3$Ru$_2$O$_7$ which also exhibits metamagnetism.~\cite{Ikeda06} This result correlates well with the large Korringa ratio~\cite{Wiecki15,Aoki19} $R_K=T_1TK_s^2/R_0$ $\approx$ 120 estimated at 20 K for UTe$_2$ (here $R_0$ is the Korringa constant of the uncorrelated system) inferred from the NMR measurements of Tokunaga {\it et al}.~\cite{Tokunaga19}.
The mechanism of the apparent change from high-temperature antiferromagnetic correlations to additional ferromagnetic fluctuations below $T^*$ can be associated with the formation of the Fermi surface of heavy quasiparticles and it is thus tempting to ascribe $T^*$ to a coherence-incoherence crossover. However, the direct observation of the ferromagnetic fluctuations by microscopic measurements is still missing in UTe$_2$. Further as shown in Fig. \ref{Fig3}, the Knight shift accurately tracks the bulk susceptiblity down to low tmperatures with no deviation below $T^*$, unlike the Knight shift anomaly observed e.g. in the heavy fermion CeMIn$_5$ (M={Co, Rh, Ir}) and AFe$_2$As$_2$ [A={K, Rb, Cs}]  where the relative weight of the local moments is continuously reduced while that of the heavy electron increases \cite{Curro2016, Shirer2012, Wu2016, Wiecki2018} upon cooling below $T^*$ .

Finally, temperature dependent magnetostriction data (derived from the difference in thermal expansion between 0 T and 10 T (see Fig.~\ref{Fig2} and Supplemental Material) allow us to calculate the $c$-axis uniaxial pressure dependence of the magnetic susceptibility for all three field directions (Supplemental Material). These data predict a suppression of $T^*$, as well as a switching of the magnetic hard direction from the $b$ axis to the $c$-axis, under $c$-axis uniaxial pressure. This is line with recent magnetization measurements under hydrostatic pressure \cite{Li2021}, probably because the largest uniaxial pressure effect (see Fig. 1a) is along the c-axis.
    
In summary, a zero-field Schottky-like anomaly at $T^*$ centered around 12 K -- observed in a variety of thermodynamic and transport probes -- governs the normal-state of UTe$_2$ above the superconducting transition.  
A high-field phase diagram of $T^*$ demonstrates that the first-order metamagnetic transition at $H_m$ follows directly from the field-evolution of $T^*$, which is also consistent with the similar Grüneisen parameters found for both $T^*$ and $H_m$. The similarity of this phase diagram with that of URhGe and our analysis of the Wilson and Korringa ratios suggests that ferromagnetic fluctuations cannot be ruled out in this system, although no direct evidence for their existence has been found to date using neutron scattering experiments \cite{Duan2020, Knafo2021}.  The magnetism in UTe2 is complicated by three different U-U distances, all near the crossover region between paramagnetic and magnetic states in the Hill plot \cite{Hill1971}.  The AFM neutron signal recently observed is consistent with AFM fluctuations between the U-U ladders along the a-axis, whereas the NMR data likely probes the U-U ladders more directly due to the close distance of the Te atom to the ladder \cite{Duan2020, Knafo2021}.  The lack of long-range magnetic order may be due to the low dimensionality of this chain-like structure, which may give a broad FM neutron scattering signal.  Clearly, more detailed experimental and theoretical work, is
needed to resolve the puzzling magnetic behavior of UTe2.

\begin{acknowledgments}
We acknowledge stimulating discussions with K. Ishida, Georg Knebel, Daniel Braithwaite, Adrien Rosuel, Jean-Pascal Brison and William Knafo.
KW acknowledges funding from the Swiss National Science Foundation through the Postdoc mobility program. Work at KIT was partially funded by the Deutsche Forschungsgemeinschaft (DFG, German Research Foundation) - TRR 288-422213477 (project A2)
\end{acknowledgments}
%-------------------------------------------------------------------------------------

\bibliographystyle{apsrev4-2}	

\begin{thebibliography}{46}%
\makeatletter
\providecommand \@ifxundefined [1]{%
 \@ifx{#1\undefined}
}%
\providecommand \@ifnum [1]{%
 \ifnum #1\expandafter \@firstoftwo
 \else \expandafter \@secondoftwo
 \fi
}%
\providecommand \@ifx [1]{%
 \ifx #1\expandafter \@firstoftwo
 \else \expandafter \@secondoftwo
 \fi
}%
\providecommand \natexlab [1]{#1}%
\providecommand \enquote  [1]{``#1''}%
\providecommand \bibnamefont  [1]{#1}%
\providecommand \bibfnamefont [1]{#1}%
\providecommand \citenamefont [1]{#1}%
\providecommand \href@noop [0]{\@secondoftwo}%
\providecommand \href [0]{\begingroup \@sanitize@url \@href}%
\providecommand \@href[1]{\@@startlink{#1}\@@href}%
\providecommand \@@href[1]{\endgroup#1\@@endlink}%
\providecommand \@sanitize@url [0]{\catcode `\\12\catcode `\$12\catcode
  `\&12\catcode `\#12\catcode `\^12\catcode `\_12\catcode `\%12\relax}%
\providecommand \@@startlink[1]{}%
\providecommand \@@endlink[0]{}%
\providecommand \url  [0]{\begingroup\@sanitize@url \@url }%
\providecommand \@url [1]{\endgroup\@href {#1}{\urlprefix }}%
\providecommand \urlprefix  [0]{URL }%
\providecommand \Eprint [0]{\href }%
\providecommand \doibase [0]{https://doi.org/}%
\providecommand \selectlanguage [0]{\@gobble}%
\providecommand \bibinfo  [0]{\@secondoftwo}%
\providecommand \bibfield  [0]{\@secondoftwo}%
\providecommand \translation [1]{[#1]}%
\providecommand \BibitemOpen [0]{}%
\providecommand \bibitemStop [0]{}%
\providecommand \bibitemNoStop [0]{.\EOS\space}%
\providecommand \EOS [0]{\spacefactor3000\relax}%
\providecommand \BibitemShut  [1]{\csname bibitem#1\endcsname}%
\let\auto@bib@innerbib\@empty
%</preamble>
\bibitem [{\citenamefont {Ran}\ \emph {et~al.}(2019{\natexlab{a}})\citenamefont
  {Ran}, \citenamefont {Eckberg}, \citenamefont {Ding}, \citenamefont
  {Furukawa}, \citenamefont {Metz}, \citenamefont {Saha}, \citenamefont {Liu},
  \citenamefont {Zic}, \citenamefont {Kim}, \citenamefont {Paglione},\ and\
  \citenamefont {Butch}}]{Ran19}%
  \BibitemOpen
  \bibfield  {author} {\bibinfo {author} {\bibfnamefont {S.}~\bibnamefont
  {Ran}}, \bibinfo {author} {\bibfnamefont {C.}~\bibnamefont {Eckberg}},
  \bibinfo {author} {\bibfnamefont {Q.-P.}\ \bibnamefont {Ding}}, \bibinfo
  {author} {\bibfnamefont {Y.}~\bibnamefont {Furukawa}}, \bibinfo {author}
  {\bibfnamefont {T.}~\bibnamefont {Metz}}, \bibinfo {author} {\bibfnamefont
  {S.~R.}\ \bibnamefont {Saha}}, \bibinfo {author} {\bibfnamefont {I.-L.}\
  \bibnamefont {Liu}}, \bibinfo {author} {\bibfnamefont {M.}~\bibnamefont
  {Zic}}, \bibinfo {author} {\bibfnamefont {H.}~\bibnamefont {Kim}}, \bibinfo
  {author} {\bibfnamefont {J.}~\bibnamefont {Paglione}},\ and\ \bibinfo
  {author} {\bibfnamefont {N.~P.}\ \bibnamefont {Butch}},\ }\href
  {https://doi.org/10.1126/science.aav8645} {\bibfield  {journal} {\bibinfo
  {journal} {Science}\ }\textbf {\bibinfo {volume} {365}},\ \bibinfo {pages}
  {684} (\bibinfo {year} {2019}{\natexlab{a}})}\BibitemShut {NoStop}%
\bibitem [{\citenamefont {Aoki}\ \emph
  {et~al.}(2019{\natexlab{a}})\citenamefont {Aoki}, \citenamefont {Nakamura},
  \citenamefont {Honda}, \citenamefont {Li}, \citenamefont {Homma},
  \citenamefont {Shimizu}, \citenamefont {Sato}, \citenamefont {Knebel},
  \citenamefont {Brison}, \citenamefont {Pourret}, \citenamefont {Braithwaite},
  \citenamefont {Lapertot}, \citenamefont {Niu}, \citenamefont {Vali{\v{s}}ka},
  \citenamefont {Harima},\ and\ \citenamefont {Flouquet}}]{Aoki19-2}%
  \BibitemOpen
  \bibfield  {author} {\bibinfo {author} {\bibfnamefont {D.}~\bibnamefont
  {Aoki}}, \bibinfo {author} {\bibfnamefont {A.}~\bibnamefont {Nakamura}},
  \bibinfo {author} {\bibfnamefont {F.}~\bibnamefont {Honda}}, \bibinfo
  {author} {\bibfnamefont {D.~X.}\ \bibnamefont {Li}}, \bibinfo {author}
  {\bibfnamefont {Y.}~\bibnamefont {Homma}}, \bibinfo {author} {\bibfnamefont
  {Y.}~\bibnamefont {Shimizu}}, \bibinfo {author} {\bibfnamefont {Y.~J.}\
  \bibnamefont {Sato}}, \bibinfo {author} {\bibfnamefont {G.}~\bibnamefont
  {Knebel}}, \bibinfo {author} {\bibfnamefont {J.~P.}\ \bibnamefont {Brison}},
  \bibinfo {author} {\bibfnamefont {A.}~\bibnamefont {Pourret}}, \bibinfo
  {author} {\bibfnamefont {D.}~\bibnamefont {Braithwaite}}, \bibinfo {author}
  {\bibfnamefont {G.}~\bibnamefont {Lapertot}}, \bibinfo {author}
  {\bibfnamefont {Q.}~\bibnamefont {Niu}}, \bibinfo {author} {\bibfnamefont
  {M.}~\bibnamefont {Vali{\v{s}}ka}}, \bibinfo {author} {\bibfnamefont
  {H.}~\bibnamefont {Harima}},\ and\ \bibinfo {author} {\bibfnamefont
  {J.}~\bibnamefont {Flouquet}},\ }\href
  {https://doi.org/10.7566/JPSJ.88.043702} {\bibfield  {journal} {\bibinfo
  {journal} {Journal of the Physical Society of Japan}\ }\textbf {\bibinfo
  {volume} {88}},\ \bibinfo {pages} {43702} (\bibinfo {year}
  {2019}{\natexlab{a}})},\ \Eprint {https://arxiv.org/abs/1903.02410}
  {arXiv:1903.02410} \BibitemShut {NoStop}%
\bibitem [{\citenamefont {Aoki}\ \emph
  {et~al.}(2019{\natexlab{b}})\citenamefont {Aoki}, \citenamefont {Ishida},\
  and\ \citenamefont {Flouquet}}]{Aoki19}%
  \BibitemOpen
  \bibfield  {author} {\bibinfo {author} {\bibfnamefont {D.}~\bibnamefont
  {Aoki}}, \bibinfo {author} {\bibfnamefont {K.}~\bibnamefont {Ishida}},\ and\
  \bibinfo {author} {\bibfnamefont {J.}~\bibnamefont {Flouquet}},\ }\href
  {https://doi.org/10.7566/JPSJ.88.022001} {\bibfield  {journal} {\bibinfo
  {journal} {Journal of the Physical Society of Japan}\ }\textbf {\bibinfo
  {volume} {88}},\ \bibinfo {pages} {22001} (\bibinfo {year}
  {2019}{\natexlab{b}})}\BibitemShut {NoStop}%
\bibitem [{\citenamefont {Miyake}\ \emph {et~al.}(2019)\citenamefont {Miyake},
  \citenamefont {Shimizu}, \citenamefont {Sato}, \citenamefont {Li},
  \citenamefont {Nakamura}, \citenamefont {Homma}, \citenamefont {Honda},
  \citenamefont {Flouquet}, \citenamefont {Masashi},\ and\ \citenamefont
  {Aoki}}]{Miyake19}%
  \BibitemOpen
  \bibfield  {author} {\bibinfo {author} {\bibfnamefont {A.}~\bibnamefont
  {Miyake}}, \bibinfo {author} {\bibfnamefont {Y.}~\bibnamefont {Shimizu}},
  \bibinfo {author} {\bibfnamefont {Y.~J.}\ \bibnamefont {Sato}}, \bibinfo
  {author} {\bibfnamefont {D.}~\bibnamefont {Li}}, \bibinfo {author}
  {\bibfnamefont {A.}~\bibnamefont {Nakamura}}, \bibinfo {author}
  {\bibfnamefont {Y.}~\bibnamefont {Homma}}, \bibinfo {author} {\bibfnamefont
  {F.}~\bibnamefont {Honda}}, \bibinfo {author} {\bibfnamefont
  {J.}~\bibnamefont {Flouquet}}, \bibinfo {author} {\bibfnamefont
  {T.}~\bibnamefont {Masashi}},\ and\ \bibinfo {author} {\bibfnamefont
  {D.}~\bibnamefont {Aoki}},\ }\href {https://doi.org/10.7566/JPSJ.88.063706}
  {\bibfield  {journal} {\bibinfo  {journal} {Journal of the Physical Society
  of Japan}\ }\textbf {\bibinfo {volume} {88}},\ \bibinfo {pages} {63706}
  (\bibinfo {year} {2019})}\BibitemShut {NoStop}%
\bibitem [{\citenamefont {William}\ \emph {et~al.}(2019)\citenamefont
  {William}, \citenamefont {Michal}, \citenamefont {Daniel}, \citenamefont
  {G{\'{e}}rard}, \citenamefont {Georg}, \citenamefont {Alexandre},
  \citenamefont {Jean-Pascal}, \citenamefont {Jacques},\ and\ \citenamefont
  {Dai}}]{Knafo19}%
  \BibitemOpen
  \bibfield  {author} {\bibinfo {author} {\bibfnamefont {K.}~\bibnamefont
  {William}}, \bibinfo {author} {\bibfnamefont {V.}~\bibnamefont {Michal}},
  \bibinfo {author} {\bibfnamefont {B.}~\bibnamefont {Daniel}}, \bibinfo
  {author} {\bibfnamefont {L.}~\bibnamefont {G{\'{e}}rard}}, \bibinfo {author}
  {\bibfnamefont {K.}~\bibnamefont {Georg}}, \bibinfo {author} {\bibfnamefont
  {P.}~\bibnamefont {Alexandre}}, \bibinfo {author} {\bibfnamefont
  {B.}~\bibnamefont {Jean-Pascal}}, \bibinfo {author} {\bibfnamefont
  {F.}~\bibnamefont {Jacques}},\ and\ \bibinfo {author} {\bibfnamefont
  {A.}~\bibnamefont {Dai}},\ }\href {https://doi.org/10.7566/JPSJ.88.063705}
  {\bibfield  {journal} {\bibinfo  {journal} {Journal of the Physical Society
  of Japan}\ }\textbf {\bibinfo {volume} {88}},\ \bibinfo {pages} {63705}
  (\bibinfo {year} {2019})}\BibitemShut {NoStop}%
\bibitem [{\citenamefont {Knebel}\ \emph {et~al.}(2019)\citenamefont {Knebel},
  \citenamefont {Knafo}, \citenamefont {Pourret}, \citenamefont {Niu},
  \citenamefont {Vali{\v{s}}ka}, \citenamefont {Braithwaite}, \citenamefont
  {Lapertot}, \citenamefont {Nardone}, \citenamefont {Zitouni}, \citenamefont
  {Mishra}, \citenamefont {Sheikin}, \citenamefont {Seyfarth}, \citenamefont
  {Brison}, \citenamefont {Aoki},\ and\ \citenamefont {Flouquet}}]{Knebel19}%
  \BibitemOpen
  \bibfield  {author} {\bibinfo {author} {\bibfnamefont {G.}~\bibnamefont
  {Knebel}}, \bibinfo {author} {\bibfnamefont {W.}~\bibnamefont {Knafo}},
  \bibinfo {author} {\bibfnamefont {A.}~\bibnamefont {Pourret}}, \bibinfo
  {author} {\bibfnamefont {Q.}~\bibnamefont {Niu}}, \bibinfo {author}
  {\bibfnamefont {M.}~\bibnamefont {Vali{\v{s}}ka}}, \bibinfo {author}
  {\bibfnamefont {D.}~\bibnamefont {Braithwaite}}, \bibinfo {author}
  {\bibfnamefont {G.}~\bibnamefont {Lapertot}}, \bibinfo {author}
  {\bibfnamefont {M.}~\bibnamefont {Nardone}}, \bibinfo {author} {\bibfnamefont
  {A.}~\bibnamefont {Zitouni}}, \bibinfo {author} {\bibfnamefont
  {S.}~\bibnamefont {Mishra}}, \bibinfo {author} {\bibfnamefont
  {I.}~\bibnamefont {Sheikin}}, \bibinfo {author} {\bibfnamefont
  {G.}~\bibnamefont {Seyfarth}}, \bibinfo {author} {\bibfnamefont {J.-P.}\
  \bibnamefont {Brison}}, \bibinfo {author} {\bibfnamefont {D.}~\bibnamefont
  {Aoki}},\ and\ \bibinfo {author} {\bibfnamefont {J.}~\bibnamefont
  {Flouquet}},\ }\href {https://doi.org/10.7566/JPSJ.88.063707} {\bibfield
  {journal} {\bibinfo  {journal} {Journal of the Physical Society of Japan}\
  }\textbf {\bibinfo {volume} {88}},\ \bibinfo {pages} {63707} (\bibinfo {year}
  {2019})}\BibitemShut {NoStop}%
\bibitem [{\citenamefont {Knebel}\ \emph {et~al.}(2020)\citenamefont {Knebel},
  \citenamefont {Kimata}, \citenamefont {Vali{\v{s}}ka}, \citenamefont {Honda},
  \citenamefont {Li}, \citenamefont {Braithwaite}, \citenamefont {Lapertot},
  \citenamefont {Knafo}, \citenamefont {Pourret}, \citenamefont {Sato},
  \citenamefont {Shimizu}, \citenamefont {Kihara}, \citenamefont {Brison},
  \citenamefont {Flouquet},\ and\ \citenamefont {Aoki}}]{Knebel2020}%
  \BibitemOpen
  \bibfield  {author} {\bibinfo {author} {\bibfnamefont {G.}~\bibnamefont
  {Knebel}}, \bibinfo {author} {\bibfnamefont {M.}~\bibnamefont {Kimata}},
  \bibinfo {author} {\bibfnamefont {M.}~\bibnamefont {Vali{\v{s}}ka}}, \bibinfo
  {author} {\bibfnamefont {F.}~\bibnamefont {Honda}}, \bibinfo {author}
  {\bibfnamefont {D.~X.}\ \bibnamefont {Li}}, \bibinfo {author} {\bibfnamefont
  {D.}~\bibnamefont {Braithwaite}}, \bibinfo {author} {\bibfnamefont
  {G.}~\bibnamefont {Lapertot}}, \bibinfo {author} {\bibfnamefont
  {W.}~\bibnamefont {Knafo}}, \bibinfo {author} {\bibfnamefont
  {A.}~\bibnamefont {Pourret}}, \bibinfo {author} {\bibfnamefont {Y.~J.}\
  \bibnamefont {Sato}}, \bibinfo {author} {\bibfnamefont {Y.}~\bibnamefont
  {Shimizu}}, \bibinfo {author} {\bibfnamefont {T.}~\bibnamefont {Kihara}},
  \bibinfo {author} {\bibfnamefont {J.~P.}\ \bibnamefont {Brison}}, \bibinfo
  {author} {\bibfnamefont {J.}~\bibnamefont {Flouquet}},\ and\ \bibinfo
  {author} {\bibfnamefont {D.}~\bibnamefont {Aoki}},\ }\href
  {https://doi.org/10.7566/JPSJ.89.053707} {\bibfield  {journal} {\bibinfo
  {journal} {Journal of the Physical Society of Japan}\ }\textbf {\bibinfo
  {volume} {89}},\ \bibinfo {pages} {1} (\bibinfo {year} {2020})},\ \Eprint
  {https://arxiv.org/abs/2003.08728} {arXiv:2003.08728} \BibitemShut {NoStop}%
\bibitem [{\citenamefont {Ran}\ \emph {et~al.}(2019{\natexlab{b}})\citenamefont
  {Ran}, \citenamefont {Liu}, \citenamefont {Eo},\ and\ \citenamefont
  {et~al.}}]{RanNatPhys}%
  \BibitemOpen
  \bibfield  {author} {\bibinfo {author} {\bibfnamefont {S.}~\bibnamefont
  {Ran}}, \bibinfo {author} {\bibfnamefont {I.}~\bibnamefont {Liu}}, \bibinfo
  {author} {\bibfnamefont {Y.}~\bibnamefont {Eo}},\ and\ \bibinfo {author}
  {\bibnamefont {et~al.}},\ }\href@noop {} {\bibfield  {journal} {\bibinfo
  {journal} {Nat. Phys}\ }\textbf {\bibinfo {volume} {15}},\ \bibinfo {pages}
  {1250} (\bibinfo {year} {2019}{\natexlab{b}})}\BibitemShut {NoStop}%
\bibitem [{\citenamefont {Aoki}\ \emph {et~al.}(2020)\citenamefont {Aoki},
  \citenamefont {Honda}, \citenamefont {Knebel}, \citenamefont {Braithwaite},
  \citenamefont {Nakamura}, \citenamefont {Li}, \citenamefont {Homma},
  \citenamefont {Shimizu}, \citenamefont {Sato}, \citenamefont {Brison},\ and\
  \citenamefont {Flouquet}}]{Aoki2020}%
  \BibitemOpen
  \bibfield  {author} {\bibinfo {author} {\bibfnamefont {D.}~\bibnamefont
  {Aoki}}, \bibinfo {author} {\bibfnamefont {F.}~\bibnamefont {Honda}},
  \bibinfo {author} {\bibfnamefont {G.}~\bibnamefont {Knebel}}, \bibinfo
  {author} {\bibfnamefont {D.}~\bibnamefont {Braithwaite}}, \bibinfo {author}
  {\bibfnamefont {A.}~\bibnamefont {Nakamura}}, \bibinfo {author}
  {\bibfnamefont {D.~X.}\ \bibnamefont {Li}}, \bibinfo {author} {\bibfnamefont
  {Y.}~\bibnamefont {Homma}}, \bibinfo {author} {\bibfnamefont
  {Y.}~\bibnamefont {Shimizu}}, \bibinfo {author} {\bibfnamefont {Y.~J.}\
  \bibnamefont {Sato}}, \bibinfo {author} {\bibfnamefont {J.~P.}\ \bibnamefont
  {Brison}},\ and\ \bibinfo {author} {\bibfnamefont {J.}~\bibnamefont
  {Flouquet}},\ }\href {https://doi.org/10.7566/JPSJ.89.053705} {\bibfield
  {journal} {\bibinfo  {journal} {Journal of the Physical Society of Japan}\
  }\textbf {\bibinfo {volume} {89}},\ \bibinfo {pages} {1} (\bibinfo {year}
  {2020})},\ \Eprint {https://arxiv.org/abs/2003.09782} {arXiv:2003.09782}
  \BibitemShut {NoStop}%
\bibitem [{\citenamefont {Braithwaite}\ \emph {et~al.}(2019)\citenamefont
  {Braithwaite}, \citenamefont {Vali{\v{s}}ka}, \citenamefont {Knebel},
  \citenamefont {Lapertot}, \citenamefont {Brison}, \citenamefont {Pourret},
  \citenamefont {Zhitomirsky}, \citenamefont {Flouquet}, \citenamefont
  {Honda},\ and\ \citenamefont {Aoki}}]{Braithwaite2019}%
  \BibitemOpen
  \bibfield  {author} {\bibinfo {author} {\bibfnamefont {D.}~\bibnamefont
  {Braithwaite}}, \bibinfo {author} {\bibfnamefont {M.}~\bibnamefont
  {Vali{\v{s}}ka}}, \bibinfo {author} {\bibfnamefont {G.}~\bibnamefont
  {Knebel}}, \bibinfo {author} {\bibfnamefont {G.}~\bibnamefont {Lapertot}},
  \bibinfo {author} {\bibfnamefont {J.~P.}\ \bibnamefont {Brison}}, \bibinfo
  {author} {\bibfnamefont {A.}~\bibnamefont {Pourret}}, \bibinfo {author}
  {\bibfnamefont {M.~E.}\ \bibnamefont {Zhitomirsky}}, \bibinfo {author}
  {\bibfnamefont {J.}~\bibnamefont {Flouquet}}, \bibinfo {author}
  {\bibfnamefont {F.}~\bibnamefont {Honda}},\ and\ \bibinfo {author}
  {\bibfnamefont {D.}~\bibnamefont {Aoki}},\ }\bibfield  {journal} {\bibinfo
  {journal} {Communications Physics}\ }\textbf {\bibinfo {volume} {2}},\ \href
  {https://doi.org/10.1038/s42005-019-0248-z} {10.1038/s42005-019-0248-z}
  (\bibinfo {year} {2019})\BibitemShut {NoStop}%
\bibitem [{\citenamefont {Thomas}\ \emph {et~al.}(2020)\citenamefont {Thomas},
  \citenamefont {Santos}, \citenamefont {Christensen}, \citenamefont {Asaba},
  \citenamefont {Ronning}, \citenamefont {Thompson}, \citenamefont {Bauer},
  \citenamefont {Fernandes}, \citenamefont {Fabbris},\ and\ \citenamefont
  {Rosa}}]{Thomas2020}%
  \BibitemOpen
  \bibfield  {author} {\bibinfo {author} {\bibfnamefont {S.~M.}\ \bibnamefont
  {Thomas}}, \bibinfo {author} {\bibfnamefont {F.~B.}\ \bibnamefont {Santos}},
  \bibinfo {author} {\bibfnamefont {M.~H.}\ \bibnamefont {Christensen}},
  \bibinfo {author} {\bibfnamefont {T.}~\bibnamefont {Asaba}}, \bibinfo
  {author} {\bibfnamefont {F.}~\bibnamefont {Ronning}}, \bibinfo {author}
  {\bibfnamefont {J.~D.}\ \bibnamefont {Thompson}}, \bibinfo {author}
  {\bibfnamefont {E.~D.}\ \bibnamefont {Bauer}}, \bibinfo {author}
  {\bibfnamefont {R.~M.}\ \bibnamefont {Fernandes}}, \bibinfo {author}
  {\bibfnamefont {G.}~\bibnamefont {Fabbris}},\ and\ \bibinfo {author}
  {\bibfnamefont {P.~F.}\ \bibnamefont {Rosa}},\ }\href@noop {} {\bibfield
  {journal} {\bibinfo  {journal} {Science Advances}\ }\textbf {\bibinfo
  {volume} {6}} (\bibinfo {year} {2020})}\BibitemShut {NoStop}%
\bibitem [{\citenamefont {Duan}\ \emph
  {et~al.}(2020{\natexlab{a}})\citenamefont {Duan}, \citenamefont {Sasmal},
  \citenamefont {Maple}, \citenamefont {Podlesnyak}, \citenamefont {Zhu},
  \citenamefont {Si},\ and\ \citenamefont {Dai}}]{Duan20}%
  \BibitemOpen
  \bibfield  {author} {\bibinfo {author} {\bibfnamefont {C.}~\bibnamefont
  {Duan}}, \bibinfo {author} {\bibfnamefont {K.}~\bibnamefont {Sasmal}},
  \bibinfo {author} {\bibfnamefont {M.~B.}\ \bibnamefont {Maple}}, \bibinfo
  {author} {\bibfnamefont {A.}~\bibnamefont {Podlesnyak}}, \bibinfo {author}
  {\bibfnamefont {J.-X.}\ \bibnamefont {Zhu}}, \bibinfo {author} {\bibfnamefont
  {Q.}~\bibnamefont {Si}},\ and\ \bibinfo {author} {\bibfnamefont
  {P.}~\bibnamefont {Dai}},\ }\href
  {https://doi.org/10.1103/PhysRevLett.125.237003} {\bibfield  {journal}
  {\bibinfo  {journal} {Phys. Rev. Lett.}\ }\textbf {\bibinfo {volume} {125}},\
  \bibinfo {pages} {237003} (\bibinfo {year} {2020}{\natexlab{a}})}\BibitemShut
  {NoStop}%
\bibitem [{\citenamefont {Miao}\ \emph {et~al.}(2020)\citenamefont {Miao},
  \citenamefont {Liu}, \citenamefont {Xu}, \citenamefont {Kotta}, \citenamefont
  {Kang}, \citenamefont {Ran}, \citenamefont {Paglione}, \citenamefont
  {Kotliar}, \citenamefont {Butch}, \citenamefont {Denlinger},\ and\
  \citenamefont {Wray}}]{Miao20}%
  \BibitemOpen
  \bibfield  {author} {\bibinfo {author} {\bibfnamefont {L.}~\bibnamefont
  {Miao}}, \bibinfo {author} {\bibfnamefont {S.}~\bibnamefont {Liu}}, \bibinfo
  {author} {\bibfnamefont {Y.}~\bibnamefont {Xu}}, \bibinfo {author}
  {\bibfnamefont {E.~C.}\ \bibnamefont {Kotta}}, \bibinfo {author}
  {\bibfnamefont {C.-J.}\ \bibnamefont {Kang}}, \bibinfo {author}
  {\bibfnamefont {S.}~\bibnamefont {Ran}}, \bibinfo {author} {\bibfnamefont
  {J.}~\bibnamefont {Paglione}}, \bibinfo {author} {\bibfnamefont
  {G.}~\bibnamefont {Kotliar}}, \bibinfo {author} {\bibfnamefont {N.~P.}\
  \bibnamefont {Butch}}, \bibinfo {author} {\bibfnamefont {J.~D.}\ \bibnamefont
  {Denlinger}},\ and\ \bibinfo {author} {\bibfnamefont {L.~A.}\ \bibnamefont
  {Wray}},\ }\href {https://doi.org/10.1103/PhysRevLett.124.076401} {\bibfield
  {journal} {\bibinfo  {journal} {Phys. Rev. Lett.}\ }\textbf {\bibinfo
  {volume} {124}},\ \bibinfo {pages} {76401} (\bibinfo {year}
  {2020})}\BibitemShut {NoStop}%
\bibitem [{\citenamefont {Tokunaga}\ \emph {et~al.}(2019)\citenamefont
  {Tokunaga}, \citenamefont {Sakai}, \citenamefont {Kambe}, \citenamefont
  {Hattori}, \citenamefont {Higa}, \citenamefont {Nakamine}, \citenamefont
  {Kitagawa}, \citenamefont {Ishida}, \citenamefont {Nakamura}, \citenamefont
  {Shimizu}, \citenamefont {Homma}, \citenamefont {Li}, \citenamefont {Honda},\
  and\ \citenamefont {Aoki}}]{Tokunaga19}%
  \BibitemOpen
  \bibfield  {author} {\bibinfo {author} {\bibfnamefont {Y.}~\bibnamefont
  {Tokunaga}}, \bibinfo {author} {\bibfnamefont {H.}~\bibnamefont {Sakai}},
  \bibinfo {author} {\bibfnamefont {S.}~\bibnamefont {Kambe}}, \bibinfo
  {author} {\bibfnamefont {T.}~\bibnamefont {Hattori}}, \bibinfo {author}
  {\bibfnamefont {N.}~\bibnamefont {Higa}}, \bibinfo {author} {\bibfnamefont
  {G.}~\bibnamefont {Nakamine}}, \bibinfo {author} {\bibfnamefont
  {S.}~\bibnamefont {Kitagawa}}, \bibinfo {author} {\bibfnamefont
  {K.}~\bibnamefont {Ishida}}, \bibinfo {author} {\bibfnamefont
  {A.}~\bibnamefont {Nakamura}}, \bibinfo {author} {\bibfnamefont
  {Y.}~\bibnamefont {Shimizu}}, \bibinfo {author} {\bibfnamefont
  {Y.}~\bibnamefont {Homma}}, \bibinfo {author} {\bibfnamefont
  {D.}~\bibnamefont {Li}}, \bibinfo {author} {\bibfnamefont {F.}~\bibnamefont
  {Honda}},\ and\ \bibinfo {author} {\bibfnamefont {D.}~\bibnamefont {Aoki}},\
  }\href {https://doi.org/10.7566/JPSJ.88.073701} {\bibfield  {journal}
  {\bibinfo  {journal} {Journal of the Physical Society of Japan}\ }\textbf
  {\bibinfo {volume} {88}},\ \bibinfo {pages} {73701} (\bibinfo {year}
  {2019})}\BibitemShut {NoStop}%
\bibitem [{\citenamefont {Sundar}\ \emph {et~al.}(2019)\citenamefont {Sundar},
  \citenamefont {Gheidi}, \citenamefont {Akintola}, \citenamefont
  {C{\^{o}}t{\'{e}}}, \citenamefont {Dunsiger}, \citenamefont {Ran},
  \citenamefont {Butch}, \citenamefont {Saha}, \citenamefont {Paglione},\ and\
  \citenamefont {Sonier}}]{Sundar19}%
  \BibitemOpen
  \bibfield  {author} {\bibinfo {author} {\bibfnamefont {S.}~\bibnamefont
  {Sundar}}, \bibinfo {author} {\bibfnamefont {S.}~\bibnamefont {Gheidi}},
  \bibinfo {author} {\bibfnamefont {K.}~\bibnamefont {Akintola}}, \bibinfo
  {author} {\bibfnamefont {A.~M.}\ \bibnamefont {C{\^{o}}t{\'{e}}}}, \bibinfo
  {author} {\bibfnamefont {S.~R.}\ \bibnamefont {Dunsiger}}, \bibinfo {author}
  {\bibfnamefont {S.}~\bibnamefont {Ran}}, \bibinfo {author} {\bibfnamefont
  {N.~P.}\ \bibnamefont {Butch}}, \bibinfo {author} {\bibfnamefont {S.~R.}\
  \bibnamefont {Saha}}, \bibinfo {author} {\bibfnamefont {J.}~\bibnamefont
  {Paglione}},\ and\ \bibinfo {author} {\bibfnamefont {J.~E.}\ \bibnamefont
  {Sonier}},\ }\href {https://doi.org/10.1103/PhysRevB.100.140502} {\bibfield
  {journal} {\bibinfo  {journal} {Phys. Rev. B}\ }\textbf {\bibinfo {volume}
  {100}},\ \bibinfo {pages} {140502} (\bibinfo {year} {2019})}\BibitemShut
  {NoStop}%
\bibitem [{\citenamefont {Taufour}\ \emph {et~al.}(2010)\citenamefont
  {Taufour}, \citenamefont {Aoki}, \citenamefont {Knebel},\ and\ \citenamefont
  {Flouquet}}]{Taufour2010}%
  \BibitemOpen
  \bibfield  {author} {\bibinfo {author} {\bibfnamefont {V.}~\bibnamefont
  {Taufour}}, \bibinfo {author} {\bibfnamefont {D.}~\bibnamefont {Aoki}},
  \bibinfo {author} {\bibfnamefont {G.}~\bibnamefont {Knebel}},\ and\ \bibinfo
  {author} {\bibfnamefont {J.}~\bibnamefont {Flouquet}},\ }\href
  {https://doi.org/10.1103/PhysRevLett.105.217201} {\bibfield  {journal}
  {\bibinfo  {journal} {Physical Review Letters}\ }\textbf {\bibinfo {volume}
  {105}},\ \bibinfo {pages} {1} (\bibinfo {year} {2010})}\BibitemShut {NoStop}%
\bibitem [{\citenamefont {Aoki}\ \emph {et~al.}(2011)\citenamefont {Aoki},
  \citenamefont {Combier}, \citenamefont {Taufour}, \citenamefont {Matsuda},
  \citenamefont {Knebel}, \citenamefont {Kotegawa},\ and\ \citenamefont
  {Flouquet}}]{Aoki2011}%
  \BibitemOpen
  \bibfield  {author} {\bibinfo {author} {\bibfnamefont {D.}~\bibnamefont
  {Aoki}}, \bibinfo {author} {\bibfnamefont {T.}~\bibnamefont {Combier}},
  \bibinfo {author} {\bibfnamefont {V.}~\bibnamefont {Taufour}}, \bibinfo
  {author} {\bibfnamefont {T.~D.}\ \bibnamefont {Matsuda}}, \bibinfo {author}
  {\bibfnamefont {G.}~\bibnamefont {Knebel}}, \bibinfo {author} {\bibfnamefont
  {H.}~\bibnamefont {Kotegawa}},\ and\ \bibinfo {author} {\bibfnamefont
  {J.}~\bibnamefont {Flouquet}},\ }\href
  {https://doi.org/10.1143/JPSJ.80.094711} {\bibfield  {journal} {\bibinfo
  {journal} {Journal of the Physical Society of Japan}\ }\textbf {\bibinfo
  {volume} {80}},\ \bibinfo {pages} {094711} (\bibinfo {year}
  {2011})}\BibitemShut {NoStop}%
\bibitem [{\citenamefont {Ran}\ \emph {et~al.}(2020)\citenamefont {Ran},
  \citenamefont {Kim}, \citenamefont {Liu}, \citenamefont {Saha}, \citenamefont
  {Hayes}, \citenamefont {Metz}, \citenamefont {Eo}, \citenamefont {Paglione},\
  and\ \citenamefont {Butch}}]{Ran20}%
  \BibitemOpen
  \bibfield  {author} {\bibinfo {author} {\bibfnamefont {S.}~\bibnamefont
  {Ran}}, \bibinfo {author} {\bibfnamefont {H.}~\bibnamefont {Kim}}, \bibinfo
  {author} {\bibfnamefont {I.-L.}\ \bibnamefont {Liu}}, \bibinfo {author}
  {\bibfnamefont {S.~R.}\ \bibnamefont {Saha}}, \bibinfo {author}
  {\bibfnamefont {I.}~\bibnamefont {Hayes}}, \bibinfo {author} {\bibfnamefont
  {T.}~\bibnamefont {Metz}}, \bibinfo {author} {\bibfnamefont {Y.~S.}\
  \bibnamefont {Eo}}, \bibinfo {author} {\bibfnamefont {J.}~\bibnamefont
  {Paglione}},\ and\ \bibinfo {author} {\bibfnamefont {N.~P.}\ \bibnamefont
  {Butch}},\ }\href {https://doi.org/10.1103/PhysRevB.101.140503} {\bibfield
  {journal} {\bibinfo  {journal} {Phys. Rev. B}\ }\textbf {\bibinfo {volume}
  {101}},\ \bibinfo {pages} {140503} (\bibinfo {year} {2020})}\BibitemShut
  {NoStop}%
\bibitem [{\citenamefont {Miyake}\ \emph {et~al.}(2009)\citenamefont {Miyake},
  \citenamefont {Aoki},\ and\ \citenamefont {Flouquet}}]{Miyake09}%
  \BibitemOpen
  \bibfield  {author} {\bibinfo {author} {\bibfnamefont {A.}~\bibnamefont
  {Miyake}}, \bibinfo {author} {\bibfnamefont {D.}~\bibnamefont {Aoki}},\ and\
  \bibinfo {author} {\bibfnamefont {J.}~\bibnamefont {Flouquet}},\ }\href
  {https://doi.org/10.1143/JPSJ.78.063703} {\bibfield  {journal} {\bibinfo
  {journal} {Journal of the Physical Society of Japan}\ }\textbf {\bibinfo
  {volume} {78}},\ \bibinfo {pages} {063703} (\bibinfo {year} {2009})},\
  \Eprint {https://arxiv.org/abs/https://doi.org/10.1143/JPSJ.78.063703}
  {https://doi.org/10.1143/JPSJ.78.063703} \BibitemShut {NoStop}%
\bibitem [{\citenamefont {Li}\ \emph {et~al.}(2021)\citenamefont {Li},
  \citenamefont {Nakamura}, \citenamefont {Honda}, \citenamefont {Sato},
  \citenamefont {Homma}, \citenamefont {Shimizu}, \citenamefont {Ishizuka},
  \citenamefont {Yanase}, \citenamefont {Knebel}, \citenamefont {Flouquet},\
  and\ \citenamefont {Aoki}}]{Li2021}%
  \BibitemOpen
  \bibfield  {author} {\bibinfo {author} {\bibfnamefont {D.}~\bibnamefont
  {Li}}, \bibinfo {author} {\bibfnamefont {A.}~\bibnamefont {Nakamura}},
  \bibinfo {author} {\bibfnamefont {F.}~\bibnamefont {Honda}}, \bibinfo
  {author} {\bibfnamefont {Y.~J.}\ \bibnamefont {Sato}}, \bibinfo {author}
  {\bibfnamefont {Y.}~\bibnamefont {Homma}}, \bibinfo {author} {\bibfnamefont
  {Y.}~\bibnamefont {Shimizu}}, \bibinfo {author} {\bibfnamefont
  {J.}~\bibnamefont {Ishizuka}}, \bibinfo {author} {\bibfnamefont
  {Y.}~\bibnamefont {Yanase}}, \bibinfo {author} {\bibfnamefont
  {G.}~\bibnamefont {Knebel}}, \bibinfo {author} {\bibfnamefont
  {J.}~\bibnamefont {Flouquet}},\ and\ \bibinfo {author} {\bibfnamefont
  {D.}~\bibnamefont {Aoki}},\ }\href@noop {} {\  (\bibinfo {year} {2021})},\
  \Eprint {https://arxiv.org/abs/2105.08593} {arXiv:2105.08593
  [cond-mat.str-el]} \BibitemShut {NoStop}%
\bibitem [{\citenamefont {Meingast}\ \emph {et~al.}(1990)\citenamefont
  {Meingast}, \citenamefont {Blank}, \citenamefont {B{\"{u}}rkle},
  \citenamefont {Obst}, \citenamefont {Wolf}, \citenamefont {W{\"{u}}hl},
  \citenamefont {Selvamanickam},\ and\ \citenamefont {Salama}}]{Meingast1990}%
  \BibitemOpen
  \bibfield  {author} {\bibinfo {author} {\bibfnamefont {C.}~\bibnamefont
  {Meingast}}, \bibinfo {author} {\bibfnamefont {B.}~\bibnamefont {Blank}},
  \bibinfo {author} {\bibfnamefont {H.}~\bibnamefont {B{\"{u}}rkle}}, \bibinfo
  {author} {\bibfnamefont {B.}~\bibnamefont {Obst}}, \bibinfo {author}
  {\bibfnamefont {T.}~\bibnamefont {Wolf}}, \bibinfo {author} {\bibfnamefont
  {H.}~\bibnamefont {W{\"{u}}hl}}, \bibinfo {author} {\bibfnamefont
  {V.}~\bibnamefont {Selvamanickam}},\ and\ \bibinfo {author} {\bibfnamefont
  {K.}~\bibnamefont {Salama}},\ }\href
  {https://doi.org/10.1103/PhysRevB.41.11299} {\bibfield  {journal} {\bibinfo
  {journal} {Phys. Rev. B}\ }\textbf {\bibinfo {volume} {41}},\ \bibinfo
  {pages} {11299} (\bibinfo {year} {1990})}\BibitemShut {NoStop}%
\bibitem [{\citenamefont {Thomas}\ \emph {et~al.}(2021)\citenamefont {Thomas},
  \citenamefont {Stevens}, \citenamefont {Santos}, \citenamefont {Fender},
  \citenamefont {Bauer}, \citenamefont {Ronning}, \citenamefont {Thompson},
  \citenamefont {Huxley},\ and\ \citenamefont {Rosa}}]{Thomas2021}%
  \BibitemOpen
  \bibfield  {author} {\bibinfo {author} {\bibfnamefont {S.~M.}\ \bibnamefont
  {Thomas}}, \bibinfo {author} {\bibfnamefont {C.}~\bibnamefont {Stevens}},
  \bibinfo {author} {\bibfnamefont {F.~B.}\ \bibnamefont {Santos}}, \bibinfo
  {author} {\bibfnamefont {S.~S.}\ \bibnamefont {Fender}}, \bibinfo {author}
  {\bibfnamefont {E.~D.}\ \bibnamefont {Bauer}}, \bibinfo {author}
  {\bibfnamefont {F.}~\bibnamefont {Ronning}}, \bibinfo {author} {\bibfnamefont
  {J.~D.}\ \bibnamefont {Thompson}}, \bibinfo {author} {\bibfnamefont
  {A.}~\bibnamefont {Huxley}},\ and\ \bibinfo {author} {\bibfnamefont
  {P.~F.~S.}\ \bibnamefont {Rosa}},\ }\href@noop {} {\bibinfo {title}
  {Competing superconducting states in ute2 revealed by thermal expansion}}
  (\bibinfo {year} {2021}),\ \Eprint {https://arxiv.org/abs/2103.09194}
  {arXiv:2103.09194 [cond-mat.str-el]} \BibitemShut {NoStop}%
\bibitem [{\citenamefont {Cairns}\ \emph {et~al.}(2020)\citenamefont {Cairns},
  \citenamefont {Stevens}, \citenamefont {O'Neill},\ and\ \citenamefont
  {Huxley}}]{Cairns2020}%
  \BibitemOpen
  \bibfield  {author} {\bibinfo {author} {\bibfnamefont {L.~P.}\ \bibnamefont
  {Cairns}}, \bibinfo {author} {\bibfnamefont {C.~R.}\ \bibnamefont {Stevens}},
  \bibinfo {author} {\bibfnamefont {C.~D.}\ \bibnamefont {O'Neill}},\ and\
  \bibinfo {author} {\bibfnamefont {A.}~\bibnamefont {Huxley}},\ }\href
  {https://doi.org/10.1088/1361-648X/ab9c5d} {\bibfield  {journal} {\bibinfo
  {journal} {Journal of Physics: Condensed Matter}\ }\textbf {\bibinfo {volume}
  {32}},\ \bibinfo {pages} {415602} (\bibinfo {year} {2020})}\BibitemShut
  {NoStop}%
\bibitem [{\citenamefont {Eo}\ \emph {et~al.}(2021)\citenamefont {Eo},
  \citenamefont {Saha}, \citenamefont {Kim}, \citenamefont {Ran}, \citenamefont
  {Horn}, \citenamefont {Hodovanets}, \citenamefont {Collini}, \citenamefont
  {Fuhrman}, \citenamefont {Nevidomskyy}, \citenamefont {Butch}, \citenamefont
  {Fuhrer},\ and\ \citenamefont {Paglione}}]{Eo2021}%
  \BibitemOpen
  \bibfield  {author} {\bibinfo {author} {\bibfnamefont {Y.~S.}\ \bibnamefont
  {Eo}}, \bibinfo {author} {\bibfnamefont {S.~R.}\ \bibnamefont {Saha}},
  \bibinfo {author} {\bibfnamefont {H.}~\bibnamefont {Kim}}, \bibinfo {author}
  {\bibfnamefont {S.}~\bibnamefont {Ran}}, \bibinfo {author} {\bibfnamefont
  {J.~A.}\ \bibnamefont {Horn}}, \bibinfo {author} {\bibfnamefont
  {H.}~\bibnamefont {Hodovanets}}, \bibinfo {author} {\bibfnamefont
  {J.}~\bibnamefont {Collini}}, \bibinfo {author} {\bibfnamefont {W.~T.}\
  \bibnamefont {Fuhrman}}, \bibinfo {author} {\bibfnamefont {A.~H.}\
  \bibnamefont {Nevidomskyy}}, \bibinfo {author} {\bibfnamefont {N.~P.}\
  \bibnamefont {Butch}}, \bibinfo {author} {\bibfnamefont {M.~S.}\ \bibnamefont
  {Fuhrer}},\ and\ \bibinfo {author} {\bibfnamefont {J.}~\bibnamefont
  {Paglione}},\ }\href@noop {} {\bibinfo {title} {Anomalous c-axis transport
  response of ute$_{2}$}} (\bibinfo {year} {2021}),\ \Eprint
  {https://arxiv.org/abs/2101.03102} {arXiv:2101.03102 [cond-mat.str-el]}
  \BibitemShut {NoStop}%
\bibitem [{\citenamefont {Niu}\ \emph {et~al.}(2020)\citenamefont {Niu},
  \citenamefont {Knebel}, \citenamefont {Braithwaite}, \citenamefont {Aoki},
  \citenamefont {Lapertot}, \citenamefont {Seyfarth}, \citenamefont {Brison},
  \citenamefont {Flouquet},\ and\ \citenamefont {Pourret}}]{Niu20}%
  \BibitemOpen
  \bibfield  {author} {\bibinfo {author} {\bibfnamefont {Q.}~\bibnamefont
  {Niu}}, \bibinfo {author} {\bibfnamefont {G.}~\bibnamefont {Knebel}},
  \bibinfo {author} {\bibfnamefont {D.}~\bibnamefont {Braithwaite}}, \bibinfo
  {author} {\bibfnamefont {D.}~\bibnamefont {Aoki}}, \bibinfo {author}
  {\bibfnamefont {G.}~\bibnamefont {Lapertot}}, \bibinfo {author}
  {\bibfnamefont {G.}~\bibnamefont {Seyfarth}}, \bibinfo {author}
  {\bibfnamefont {J.-P.}\ \bibnamefont {Brison}}, \bibinfo {author}
  {\bibfnamefont {J.}~\bibnamefont {Flouquet}},\ and\ \bibinfo {author}
  {\bibfnamefont {A.}~\bibnamefont {Pourret}},\ }\href
  {https://doi.org/10.1103/PhysRevLett.124.086601} {\bibfield  {journal}
  {\bibinfo  {journal} {Phys. Rev. Lett.}\ }\textbf {\bibinfo {volume} {124}},\
  \bibinfo {pages} {86601} (\bibinfo {year} {2020})}\BibitemShut {NoStop}%
\bibitem [{\citenamefont {Imajo}\ \emph {et~al.}(2019)\citenamefont {Imajo},
  \citenamefont {Kohama}, \citenamefont {Miyake}, \citenamefont {Dong},
  \citenamefont {Tokunaga}, \citenamefont {Flouquet}, \citenamefont {Kindo},\
  and\ \citenamefont {Aoki}}]{Imajo19}%
  \BibitemOpen
  \bibfield  {author} {\bibinfo {author} {\bibfnamefont {S.}~\bibnamefont
  {Imajo}}, \bibinfo {author} {\bibfnamefont {Y.}~\bibnamefont {Kohama}},
  \bibinfo {author} {\bibfnamefont {A.}~\bibnamefont {Miyake}}, \bibinfo
  {author} {\bibfnamefont {C.}~\bibnamefont {Dong}}, \bibinfo {author}
  {\bibfnamefont {M.}~\bibnamefont {Tokunaga}}, \bibinfo {author}
  {\bibfnamefont {J.}~\bibnamefont {Flouquet}}, \bibinfo {author}
  {\bibfnamefont {K.}~\bibnamefont {Kindo}},\ and\ \bibinfo {author}
  {\bibfnamefont {D.}~\bibnamefont {Aoki}},\ }\href
  {https://doi.org/10.7566/JPSJ.88.083705} {\bibfield  {journal} {\bibinfo
  {journal} {Journal of the Physical Society of Japan}\ }\textbf {\bibinfo
  {volume} {88}},\ \bibinfo {pages} {83705} (\bibinfo {year}
  {2019})}\BibitemShut {NoStop}%
\bibitem [{\citenamefont {Fisher}\ and\ \citenamefont
  {Langer}(1968)}]{Fisher68}%
  \BibitemOpen
  \bibfield  {author} {\bibinfo {author} {\bibfnamefont {M.~E.}\ \bibnamefont
  {Fisher}}\ and\ \bibinfo {author} {\bibfnamefont {J.~S.}\ \bibnamefont
  {Langer}},\ }\href {https://doi.org/10.1103/PhysRevLett.20.665} {\bibfield
  {journal} {\bibinfo  {journal} {Phys. Rev. Lett.}\ }\textbf {\bibinfo
  {volume} {20}},\ \bibinfo {pages} {665} (\bibinfo {year} {1968})}\BibitemShut
  {NoStop}%
\bibitem [{\citenamefont {Geldart}\ and\ \citenamefont
  {Richard}(1975)}]{Geldart75}%
  \BibitemOpen
  \bibfield  {author} {\bibinfo {author} {\bibfnamefont {D.~J.~W.}\
  \bibnamefont {Geldart}}\ and\ \bibinfo {author} {\bibfnamefont {T.~G.}\
  \bibnamefont {Richard}},\ }\href {https://doi.org/10.1103/PhysRevB.12.5175}
  {\bibfield  {journal} {\bibinfo  {journal} {Phys. Rev. B}\ }\textbf {\bibinfo
  {volume} {12}},\ \bibinfo {pages} {5175} (\bibinfo {year}
  {1975})}\BibitemShut {NoStop}%
\bibitem [{\citenamefont {Meingast}\ \emph {et~al.}(2009)\citenamefont
  {Meingast}, \citenamefont {Zhang}, \citenamefont {Wolf}, \citenamefont
  {Hardy}, \citenamefont {Grube}, \citenamefont {Knafo}, \citenamefont
  {Adelmann}, \citenamefont {Schweiss},\ and\ \citenamefont
  {L{\"o}hneysen}}]{Meingast2009}%
  \BibitemOpen
  \bibfield  {author} {\bibinfo {author} {\bibfnamefont {C.}~\bibnamefont
  {Meingast}}, \bibinfo {author} {\bibfnamefont {Q.}~\bibnamefont {Zhang}},
  \bibinfo {author} {\bibfnamefont {T.}~\bibnamefont {Wolf}}, \bibinfo {author}
  {\bibfnamefont {F.}~\bibnamefont {Hardy}}, \bibinfo {author} {\bibfnamefont
  {K.}~\bibnamefont {Grube}}, \bibinfo {author} {\bibfnamefont
  {W.}~\bibnamefont {Knafo}}, \bibinfo {author} {\bibfnamefont
  {P.}~\bibnamefont {Adelmann}}, \bibinfo {author} {\bibfnamefont
  {P.}~\bibnamefont {Schweiss}},\ and\ \bibinfo {author} {\bibfnamefont
  {H.~v.}\ \bibnamefont {L{\"o}hneysen}},\ }in\ \href@noop {} {\emph {\bibinfo
  {booktitle} {Properties and Applications of Thermoelectric Materials}}},\
  \bibinfo {editor} {edited by\ \bibinfo {editor} {\bibfnamefont
  {V.}~\bibnamefont {Zlati{\'{c}}}}\ and\ \bibinfo {editor} {\bibfnamefont
  {A.~C.}\ \bibnamefont {Hewson}}}\ (\bibinfo  {publisher} {Springer
  Netherlands},\ \bibinfo {address} {Dordrecht},\ \bibinfo {year} {2009})\ pp.\
  \bibinfo {pages} {261--266}\BibitemShut {NoStop}%
\bibitem [{\citenamefont {Hardy}\ \emph {et~al.}(2016)\citenamefont {Hardy},
  \citenamefont {B{\"{o}}hmer}, \citenamefont {De'Medici}, \citenamefont
  {Capone}, \citenamefont {Giovannetti}, \citenamefont {Eder}, \citenamefont
  {Wang}, \citenamefont {He}, \citenamefont {Wolf}, \citenamefont {Schweiss},
  \citenamefont {Heid}, \citenamefont {Herbig}, \citenamefont {Adelmann},
  \citenamefont {Fisher},\ and\ \citenamefont {Meingast}}]{Hardy2016}%
  \BibitemOpen
  \bibfield  {author} {\bibinfo {author} {\bibfnamefont {F.}~\bibnamefont
  {Hardy}}, \bibinfo {author} {\bibfnamefont {A.~E.}\ \bibnamefont
  {B{\"{o}}hmer}}, \bibinfo {author} {\bibfnamefont {L.}~\bibnamefont
  {De'Medici}}, \bibinfo {author} {\bibfnamefont {M.}~\bibnamefont {Capone}},
  \bibinfo {author} {\bibfnamefont {G.}~\bibnamefont {Giovannetti}}, \bibinfo
  {author} {\bibfnamefont {R.}~\bibnamefont {Eder}}, \bibinfo {author}
  {\bibfnamefont {L.}~\bibnamefont {Wang}}, \bibinfo {author} {\bibfnamefont
  {M.}~\bibnamefont {He}}, \bibinfo {author} {\bibfnamefont {T.}~\bibnamefont
  {Wolf}}, \bibinfo {author} {\bibfnamefont {P.}~\bibnamefont {Schweiss}},
  \bibinfo {author} {\bibfnamefont {R.}~\bibnamefont {Heid}}, \bibinfo {author}
  {\bibfnamefont {A.}~\bibnamefont {Herbig}}, \bibinfo {author} {\bibfnamefont
  {P.}~\bibnamefont {Adelmann}}, \bibinfo {author} {\bibfnamefont {R.~A.}\
  \bibnamefont {Fisher}},\ and\ \bibinfo {author} {\bibfnamefont
  {C.}~\bibnamefont {Meingast}},\ }\href
  {https://doi.org/10.1103/PhysRevB.94.205113} {\bibfield  {journal} {\bibinfo
  {journal} {Physical Review B}\ }\textbf {\bibinfo {volume} {94}},\ \bibinfo
  {pages} {1} (\bibinfo {year} {2016})},\ \Eprint
  {https://arxiv.org/abs/1605.05485} {arXiv:1605.05485} \BibitemShut {NoStop}%
\bibitem [{\citenamefont {Hardy}\ \emph {et~al.}(2013)\citenamefont {Hardy},
  \citenamefont {B{\"{o}}hmer}, \citenamefont {Aoki}, \citenamefont {Burger},
  \citenamefont {Wolf}, \citenamefont {Schweiss}, \citenamefont {Heid},
  \citenamefont {Adelmann}, \citenamefont {Yao}, \citenamefont {Kotliar},
  \citenamefont {Schmalian},\ and\ \citenamefont {Meingast}}]{Hardy2013}%
  \BibitemOpen
  \bibfield  {author} {\bibinfo {author} {\bibfnamefont {F.}~\bibnamefont
  {Hardy}}, \bibinfo {author} {\bibfnamefont {A.~E.}\ \bibnamefont
  {B{\"{o}}hmer}}, \bibinfo {author} {\bibfnamefont {D.}~\bibnamefont {Aoki}},
  \bibinfo {author} {\bibfnamefont {P.}~\bibnamefont {Burger}}, \bibinfo
  {author} {\bibfnamefont {T.}~\bibnamefont {Wolf}}, \bibinfo {author}
  {\bibfnamefont {P.}~\bibnamefont {Schweiss}}, \bibinfo {author}
  {\bibfnamefont {R.}~\bibnamefont {Heid}}, \bibinfo {author} {\bibfnamefont
  {P.}~\bibnamefont {Adelmann}}, \bibinfo {author} {\bibfnamefont {Y.~X.}\
  \bibnamefont {Yao}}, \bibinfo {author} {\bibfnamefont {G.}~\bibnamefont
  {Kotliar}}, \bibinfo {author} {\bibfnamefont {J.}~\bibnamefont {Schmalian}},\
  and\ \bibinfo {author} {\bibfnamefont {C.}~\bibnamefont {Meingast}},\ }\href
  {https://doi.org/10.1103/PhysRevLett.111.027002} {\bibfield  {journal}
  {\bibinfo  {journal} {Physical Review Letters}\ }\textbf {\bibinfo {volume}
  {111}},\ \bibinfo {pages} {1} (\bibinfo {year} {2013})}\BibitemShut {NoStop}%
\bibitem [{\citenamefont {Wiecki}\ \emph {et~al.}(2020)\citenamefont {Wiecki},
  \citenamefont {Haghighirad}, \citenamefont {Weber}, \citenamefont {Merz},
  \citenamefont {Heid},\ and\ \citenamefont {B{\"{o}}hmer}}]{Wiecki2020}%
  \BibitemOpen
  \bibfield  {author} {\bibinfo {author} {\bibfnamefont {P.}~\bibnamefont
  {Wiecki}}, \bibinfo {author} {\bibfnamefont {A.~A.}\ \bibnamefont
  {Haghighirad}}, \bibinfo {author} {\bibfnamefont {F.}~\bibnamefont {Weber}},
  \bibinfo {author} {\bibfnamefont {M.}~\bibnamefont {Merz}}, \bibinfo {author}
  {\bibfnamefont {R.}~\bibnamefont {Heid}},\ and\ \bibinfo {author}
  {\bibfnamefont {A.~E.}\ \bibnamefont {B{\"{o}}hmer}},\ }\href
  {https://doi.org/10.1103/PhysRevLett.125.187001} {\bibfield  {journal}
  {\bibinfo  {journal} {Physical Review Letters}\ }\textbf {\bibinfo {volume}
  {125}},\ \bibinfo {pages} {1} (\bibinfo {year} {2020})},\ \Eprint
  {https://arxiv.org/abs/2005.13838} {arXiv:2005.13838} \BibitemShut {NoStop}%
\bibitem [{\citenamefont {Wiecki}\ \emph {et~al.}(2021)\citenamefont {Wiecki},
  \citenamefont {Frachet}, \citenamefont {Haghighirad}, \citenamefont {Wolf},
  \citenamefont {Meingast}, \citenamefont {Heid},\ and\ \citenamefont
  {B{\"{o}}hmer}}]{Wiecki2021}%
  \BibitemOpen
  \bibfield  {author} {\bibinfo {author} {\bibfnamefont {P.}~\bibnamefont
  {Wiecki}}, \bibinfo {author} {\bibfnamefont {M.}~\bibnamefont {Frachet}},
  \bibinfo {author} {\bibfnamefont {A.~A.}\ \bibnamefont {Haghighirad}},
  \bibinfo {author} {\bibfnamefont {T.}~\bibnamefont {Wolf}}, \bibinfo {author}
  {\bibfnamefont {C.}~\bibnamefont {Meingast}}, \bibinfo {author}
  {\bibfnamefont {R.}~\bibnamefont {Heid}},\ and\ \bibinfo {author}
  {\bibfnamefont {A.~E.}\ \bibnamefont {B{\"{o}}hmer}},\ }\href
  {http://arxiv.org/abs/2103.08972} {\  (\bibinfo {year} {2021})},\ \Eprint
  {https://arxiv.org/abs/2103.08972} {arXiv:2103.08972} \BibitemShut {NoStop}%
\bibitem [{\citenamefont {Flouquet}(2005)}]{Flouquet05}%
  \BibitemOpen
  \bibfield  {author} {\bibinfo {author} {\bibfnamefont {J.}~\bibnamefont
  {Flouquet}},\ }\href {https://doi.org/10.1016/S0079-6417(05)15002-1}
  {\bibfield  {journal} {\bibinfo  {journal} {Progress in Low Temperature
  Physics}\ }\textbf {\bibinfo {volume} {15}},\ \bibinfo {pages} {139}
  (\bibinfo {year} {2005})}\BibitemShut {NoStop}%
\bibitem [{\citenamefont {Garst}\ and\ \citenamefont
  {Rosch}(2005)}]{Garst2005}%
  \BibitemOpen
  \bibfield  {author} {\bibinfo {author} {\bibfnamefont {M.}~\bibnamefont
  {Garst}}\ and\ \bibinfo {author} {\bibfnamefont {A.}~\bibnamefont {Rosch}},\
  }\href {https://doi.org/10.1103/PhysRevB.72.205129} {\bibfield  {journal}
  {\bibinfo  {journal} {Physical Review B - Condensed Matter and Materials
  Physics}\ }\textbf {\bibinfo {volume} {72}},\ \bibinfo {pages} {1} (\bibinfo
  {year} {2005})}\BibitemShut {NoStop}%
\bibitem [{\citenamefont {Zhu}\ \emph {et~al.}(2003)\citenamefont {Zhu},
  \citenamefont {Garst}, \citenamefont {Rosch},\ and\ \citenamefont
  {Si}}]{Zhu2003}%
  \BibitemOpen
  \bibfield  {author} {\bibinfo {author} {\bibfnamefont {L.}~\bibnamefont
  {Zhu}}, \bibinfo {author} {\bibfnamefont {M.}~\bibnamefont {Garst}}, \bibinfo
  {author} {\bibfnamefont {A.}~\bibnamefont {Rosch}},\ and\ \bibinfo {author}
  {\bibfnamefont {Q.}~\bibnamefont {Si}},\ }\href
  {https://doi.org/10.1103/PhysRevLett.91.066404} {\bibfield  {journal}
  {\bibinfo  {journal} {Physical Review Letters}\ }\textbf {\bibinfo {volume}
  {91}},\ \bibinfo {pages} {6} (\bibinfo {year} {2003})},\ \Eprint
  {https://arxiv.org/abs/0212335} {arXiv:0212335 [cond-mat]} \BibitemShut
  {NoStop}%
\bibitem [{\citenamefont {Honda}\ and\ \citenamefont {et~al.}()}]{Honda}%
  \BibitemOpen
  \bibfield  {author} {\bibinfo {author} {\bibfnamefont {F.}~\bibnamefont
  {Honda}}\ and\ \bibinfo {author} {\bibnamefont {et~al.}},\ }\href@noop {}
  {\bibfield  {journal} {\bibinfo  {journal} {to be published}\ }\textbf
  {\bibinfo {volume} {??}},\ \bibinfo {pages} {????} (\bibinfo {year}
  {????})}\BibitemShut {NoStop}%
\bibitem [{\citenamefont {Ikeda}\ \emph {et~al.}(2006)\citenamefont {Ikeda},
  \citenamefont {Sakai}, \citenamefont {Aoki}, \citenamefont {Homma},
  \citenamefont {Yamamoto}, \citenamefont {Nakamura}, \citenamefont {Shiokawa},
  \citenamefont {Haga},\ and\ \citenamefont {\=Onuki}}]{Ikeda06}%
  \BibitemOpen
  \bibfield  {author} {\bibinfo {author} {\bibfnamefont {S.}~\bibnamefont
  {Ikeda}}, \bibinfo {author} {\bibfnamefont {H.}~\bibnamefont {Sakai}},
  \bibinfo {author} {\bibfnamefont {D.}~\bibnamefont {Aoki}}, \bibinfo {author}
  {\bibfnamefont {Y.}~\bibnamefont {Homma}}, \bibinfo {author} {\bibfnamefont
  {E.}~\bibnamefont {Yamamoto}}, \bibinfo {author} {\bibfnamefont
  {A.}~\bibnamefont {Nakamura}}, \bibinfo {author} {\bibfnamefont
  {Y.}~\bibnamefont {Shiokawa}}, \bibinfo {author} {\bibfnamefont
  {Y.}~\bibnamefont {Haga}},\ and\ \bibinfo {author} {\bibfnamefont
  {Y.}~\bibnamefont {\=Onuki}},\ }\href {https://doi.org/10.1143/JPSJS.75S.116}
  {\bibfield  {journal} {\bibinfo  {journal} {Journal of the Physical Society
  of Japan}\ }\textbf {\bibinfo {volume} {75}},\ \bibinfo {pages} {116}
  (\bibinfo {year} {2006})}\BibitemShut {NoStop}%
\bibitem [{\citenamefont {Wiecki}\ \emph {et~al.}(2015)\citenamefont {Wiecki},
  \citenamefont {Roy}, \citenamefont {Johnston}, \citenamefont {Bud'ko},
  \citenamefont {Canfield},\ and\ \citenamefont {Furukawa}}]{Wiecki15}%
  \BibitemOpen
  \bibfield  {author} {\bibinfo {author} {\bibfnamefont {P.}~\bibnamefont
  {Wiecki}}, \bibinfo {author} {\bibfnamefont {B.}~\bibnamefont {Roy}},
  \bibinfo {author} {\bibfnamefont {D.~C.}\ \bibnamefont {Johnston}}, \bibinfo
  {author} {\bibfnamefont {S.~L.}\ \bibnamefont {Bud'ko}}, \bibinfo {author}
  {\bibfnamefont {P.~C.}\ \bibnamefont {Canfield}},\ and\ \bibinfo {author}
  {\bibfnamefont {Y.}~\bibnamefont {Furukawa}},\ }\href
  {https://doi.org/10.1103/PhysRevLett.115.137001} {\bibfield  {journal}
  {\bibinfo  {journal} {Phys. Rev. Lett.}\ }\textbf {\bibinfo {volume} {115}},\
  \bibinfo {pages} {137001} (\bibinfo {year} {2015})}\BibitemShut {NoStop}%
\bibitem [{\citenamefont {Curro}(2016)}]{Curro2016}%
  \BibitemOpen
  \bibfield  {author} {\bibinfo {author} {\bibfnamefont {N.~J.}\ \bibnamefont
  {Curro}},\ }\href {https://doi.org/10.1088/0034-4885/79/6/064501} {\bibfield
  {journal} {\bibinfo  {journal} {Reports on Progress in Physics}\ }\textbf
  {\bibinfo {volume} {79}},\ \bibinfo {pages} {064501} (\bibinfo {year}
  {2016})}\BibitemShut {NoStop}%
\bibitem [{\citenamefont {Shirer}\ \emph {et~al.}(2012)\citenamefont {Shirer},
  \citenamefont {Shockley}, \citenamefont {Dioguardi}, \citenamefont {Crocker},
  \citenamefont {Lin}, \citenamefont {apRoberts Warren}, \citenamefont
  {Nisson}, \citenamefont {Klavins}, \citenamefont {Cooley}, \citenamefont
  {Yang},\ and\ \citenamefont {Curro}}]{Shirer2012}%
  \BibitemOpen
  \bibfield  {author} {\bibinfo {author} {\bibfnamefont {K.~R.}\ \bibnamefont
  {Shirer}}, \bibinfo {author} {\bibfnamefont {A.~C.}\ \bibnamefont
  {Shockley}}, \bibinfo {author} {\bibfnamefont {A.~P.}\ \bibnamefont
  {Dioguardi}}, \bibinfo {author} {\bibfnamefont {J.}~\bibnamefont {Crocker}},
  \bibinfo {author} {\bibfnamefont {C.~H.}\ \bibnamefont {Lin}}, \bibinfo
  {author} {\bibfnamefont {N.}~\bibnamefont {apRoberts Warren}}, \bibinfo
  {author} {\bibfnamefont {D.~M.}\ \bibnamefont {Nisson}}, \bibinfo {author}
  {\bibfnamefont {P.}~\bibnamefont {Klavins}}, \bibinfo {author} {\bibfnamefont
  {J.~C.}\ \bibnamefont {Cooley}}, \bibinfo {author} {\bibfnamefont {Y.-f.}\
  \bibnamefont {Yang}},\ and\ \bibinfo {author} {\bibfnamefont {N.~J.}\
  \bibnamefont {Curro}},\ }\href {https://doi.org/10.1073/pnas.1209609109}
  {\bibfield  {journal} {\bibinfo  {journal} {Proceedings of the National
  Academy of Sciences}\ }\textbf {\bibinfo {volume} {109}},\ \bibinfo {pages}
  {E3067} (\bibinfo {year} {2012})},\ \Eprint
  {https://arxiv.org/abs/https://www.pnas.org/content/109/45/E3067.full.pdf}
  {https://www.pnas.org/content/109/45/E3067.full.pdf} \BibitemShut {NoStop}%
\bibitem [{\citenamefont {Wu}\ \emph {et~al.}(2016)\citenamefont {Wu},
  \citenamefont {Zhao}, \citenamefont {Wang}, \citenamefont {Wang},
  \citenamefont {Xiang}, \citenamefont {Luo}, \citenamefont {Wu},\ and\
  \citenamefont {Chen}}]{Wu2016}%
  \BibitemOpen
  \bibfield  {author} {\bibinfo {author} {\bibfnamefont {Y.~P.}\ \bibnamefont
  {Wu}}, \bibinfo {author} {\bibfnamefont {D.}~\bibnamefont {Zhao}}, \bibinfo
  {author} {\bibfnamefont {A.~F.}\ \bibnamefont {Wang}}, \bibinfo {author}
  {\bibfnamefont {N.~Z.}\ \bibnamefont {Wang}}, \bibinfo {author}
  {\bibfnamefont {Z.~J.}\ \bibnamefont {Xiang}}, \bibinfo {author}
  {\bibfnamefont {X.~G.}\ \bibnamefont {Luo}}, \bibinfo {author} {\bibfnamefont
  {T.}~\bibnamefont {Wu}},\ and\ \bibinfo {author} {\bibfnamefont {X.~H.}\
  \bibnamefont {Chen}},\ }\href
  {https://doi.org/10.1103/PhysRevLett.116.147001} {\bibfield  {journal}
  {\bibinfo  {journal} {Physical Review Letters}\ }\textbf {\bibinfo {volume}
  {116}},\ \bibinfo {pages} {1} (\bibinfo {year} {2016})}\BibitemShut {NoStop}%
\bibitem [{\citenamefont {Wiecki}\ \emph {et~al.}(2018)\citenamefont {Wiecki},
  \citenamefont {Taufour}, \citenamefont {Chung}, \citenamefont {Kanatzidis},
  \citenamefont {Bud'Ko}, \citenamefont {Canfield},\ and\ \citenamefont
  {Furukawa}}]{Wiecki2018}%
  \BibitemOpen
  \bibfield  {author} {\bibinfo {author} {\bibfnamefont {P.}~\bibnamefont
  {Wiecki}}, \bibinfo {author} {\bibfnamefont {V.}~\bibnamefont {Taufour}},
  \bibinfo {author} {\bibfnamefont {D.~Y.}\ \bibnamefont {Chung}}, \bibinfo
  {author} {\bibfnamefont {M.~G.}\ \bibnamefont {Kanatzidis}}, \bibinfo
  {author} {\bibfnamefont {S.~L.}\ \bibnamefont {Bud'Ko}}, \bibinfo {author}
  {\bibfnamefont {P.~C.}\ \bibnamefont {Canfield}},\ and\ \bibinfo {author}
  {\bibfnamefont {Y.}~\bibnamefont {Furukawa}},\ }\href
  {https://doi.org/10.1103/PhysRevB.97.064509} {\bibfield  {journal} {\bibinfo
  {journal} {Physical Review B}\ }\textbf {\bibinfo {volume} {97}},\ \bibinfo
  {pages} {1} (\bibinfo {year} {2018})},\ \Eprint
  {https://arxiv.org/abs/1802.02269} {1802.02269} \BibitemShut {NoStop}%
\bibitem [{\citenamefont {Duan}\ \emph
  {et~al.}(2020{\natexlab{b}})\citenamefont {Duan}, \citenamefont {Sasmal},
  \citenamefont {Maple}, \citenamefont {Podlesnyak}, \citenamefont {Zhu},
  \citenamefont {Si},\ and\ \citenamefont {Dai}}]{Duan2020}%
  \BibitemOpen
  \bibfield  {author} {\bibinfo {author} {\bibfnamefont {C.}~\bibnamefont
  {Duan}}, \bibinfo {author} {\bibfnamefont {K.}~\bibnamefont {Sasmal}},
  \bibinfo {author} {\bibfnamefont {M.~B.}\ \bibnamefont {Maple}}, \bibinfo
  {author} {\bibfnamefont {A.}~\bibnamefont {Podlesnyak}}, \bibinfo {author}
  {\bibfnamefont {J.-X.}\ \bibnamefont {Zhu}}, \bibinfo {author} {\bibfnamefont
  {Q.}~\bibnamefont {Si}},\ and\ \bibinfo {author} {\bibfnamefont
  {P.}~\bibnamefont {Dai}},\ }\href
  {https://doi.org/10.1103/PhysRevLett.125.237003} {\bibfield  {journal}
  {\bibinfo  {journal} {Phys. Rev. Lett.}\ }\textbf {\bibinfo {volume} {125}},\
  \bibinfo {pages} {237003} (\bibinfo {year} {2020}{\natexlab{b}})}\BibitemShut
  {NoStop}%
\bibitem [{\citenamefont {Knafo}\ \emph {et~al.}(2021)\citenamefont {Knafo},
  \citenamefont {Knebel}, \citenamefont {Steffens}, \citenamefont {Kaneko},
  \citenamefont {Rosuel}, \citenamefont {Brison}, \citenamefont {Flouquet},
  \citenamefont {Aoki}, \citenamefont {Lapertot},\ and\ \citenamefont
  {Raymond}}]{Knafo2021}%
  \BibitemOpen
  \bibfield  {author} {\bibinfo {author} {\bibfnamefont {W.}~\bibnamefont
  {Knafo}}, \bibinfo {author} {\bibfnamefont {G.}~\bibnamefont {Knebel}},
  \bibinfo {author} {\bibfnamefont {P.}~\bibnamefont {Steffens}}, \bibinfo
  {author} {\bibfnamefont {K.}~\bibnamefont {Kaneko}}, \bibinfo {author}
  {\bibfnamefont {A.}~\bibnamefont {Rosuel}}, \bibinfo {author} {\bibfnamefont
  {J.~P.}\ \bibnamefont {Brison}}, \bibinfo {author} {\bibfnamefont
  {J.}~\bibnamefont {Flouquet}}, \bibinfo {author} {\bibfnamefont
  {D.}~\bibnamefont {Aoki}}, \bibinfo {author} {\bibfnamefont {G.}~\bibnamefont
  {Lapertot}},\ and\ \bibinfo {author} {\bibfnamefont {S.}~\bibnamefont
  {Raymond}},\ }\href@noop {} {\bibinfo {title} {Low-dimensional
  antiferromagnetic fluctuations in the heavy-fermion paramagnetic ladder
  ute$_2$}} (\bibinfo {year} {2021}),\ \Eprint
  {https://arxiv.org/abs/2106.13087} {arXiv:2106.13087 [cond-mat.str-el]}
  \BibitemShut {NoStop}%
\bibitem [{\citenamefont {Hill}(1971)}]{Hill1971}%
  \BibitemOpen
  \bibfield  {author} {\bibinfo {author} {\bibfnamefont {H.~H.}\ \bibnamefont
  {Hill}},\ }\bibfield  {journal} {\bibinfo  {journal} {PLUTONIUM 1970 AND
  OTHER ACTINIDES. NUCLEAR METALLURGY, VOLUME 17. Proceedings of the 4th
  International Conference, Sante Fe, New Mexico, October 5--9, 1970 (Miner, W
  N)}\ }\href {https://www.osti.gov/biblio/4080663} {} (\bibinfo {year}
  {1971})\BibitemShut {NoStop}%
\end{thebibliography}


% Encoding: UTF-8

\end{document}

\title{Thermodynamic signatures of short-range magnetic correlations in UTe$_2$}

\author{Kristin~Willa}
\affiliation{Institute for Quantum Materials and Technologies, Karlsruhe Institute of Technology, 76021 Karlsruhe, Germany}

\author{Frédéric~Hardy}
\affiliation{Institute for Quantum Materials and Technologies, Karlsruhe Institute of Technology, 76021 Karlsruhe, Germany}

\author{Dai~Aoki}
\affiliation{Institute for Materials Research, Tohoku University, Ibaraki 311-1313, Japan}
\affiliation{Univ. Grenoble Alpes, CEA, Grenoble INP, IRIG, PHELIQS, F-38000 Grenoble, France}

\author{Dexin~Li}
\affiliation{Institute for Materials Research, Tohoku University, Ibaraki 311-1313, Japan}

\author{Paul~Wiecki}
\affiliation{Institute for Quantum Materials and Technologies, Karlsruhe Institute of Technology, 76021 Karlsruhe, Germany}

\author{Gérard~Lapertot}
\affiliation{Univ. Grenoble Alpes, CEA, Grenoble INP, IRIG, PHELIQS, F-38000 Grenoble, France}

\author{Christoph~Meingast}
\affiliation{Institute for Quantum Materials and Technologies, Karlsruhe Institute of Technology, 76021 Karlsruhe, Germany}

%-------------------------------------------------------------------------------------

\maketitle

\begin{figure}[h!]
\centering
%-----------------------------------------------------------
\includegraphics[width=0.9\linewidth]{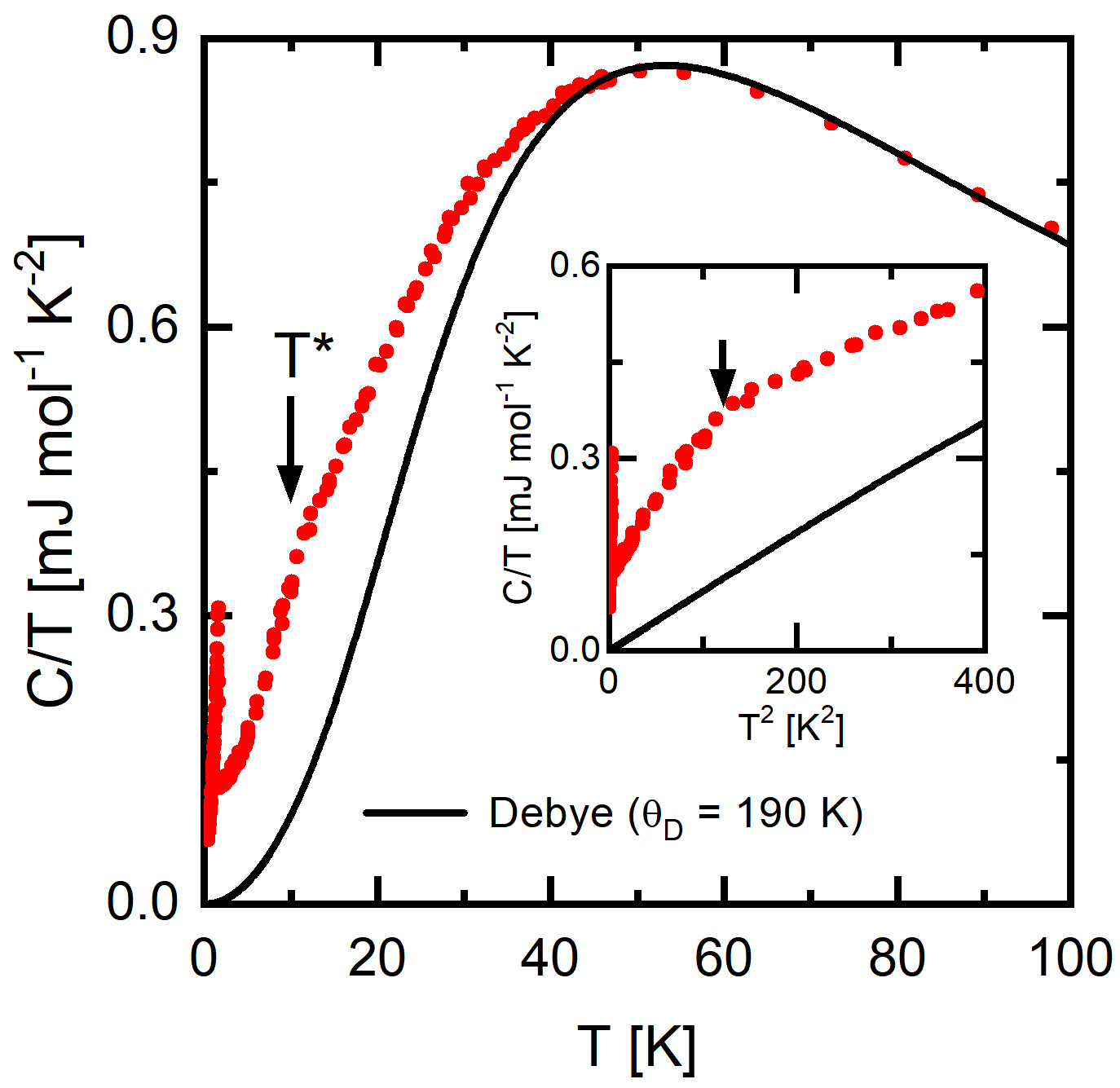}
%-----------------------------------------------------------
\caption{Specific heat over temperature vs temperature. The lattice contribution is fitted with a Debye-function. In the inset $C/T$ against $T^2$ clearly displays a change of slope around $T^*$.}      
\end{figure}

In Figure S1, we show the temperature dependence of the measured heat capacity along with a Debye specific heat with $\Theta_D=190K$. Clearly, an excess (non-phonon) heat capacity contribution is observed around $T^*$, the temperature where the thermal expansion exhibits a deep minimum. This is also clearly seen as a large change of slope in the inset of S1(b), where $C/T$ is plotted as a function of $T^2$. The resulting electronic/magnetic contribution to the specific heat is displayed in Fig. 1 in the main text. 
\begin{figure}[t!]
\centering
%-----------------------------------------------------------
\includegraphics[width=0.9\linewidth]{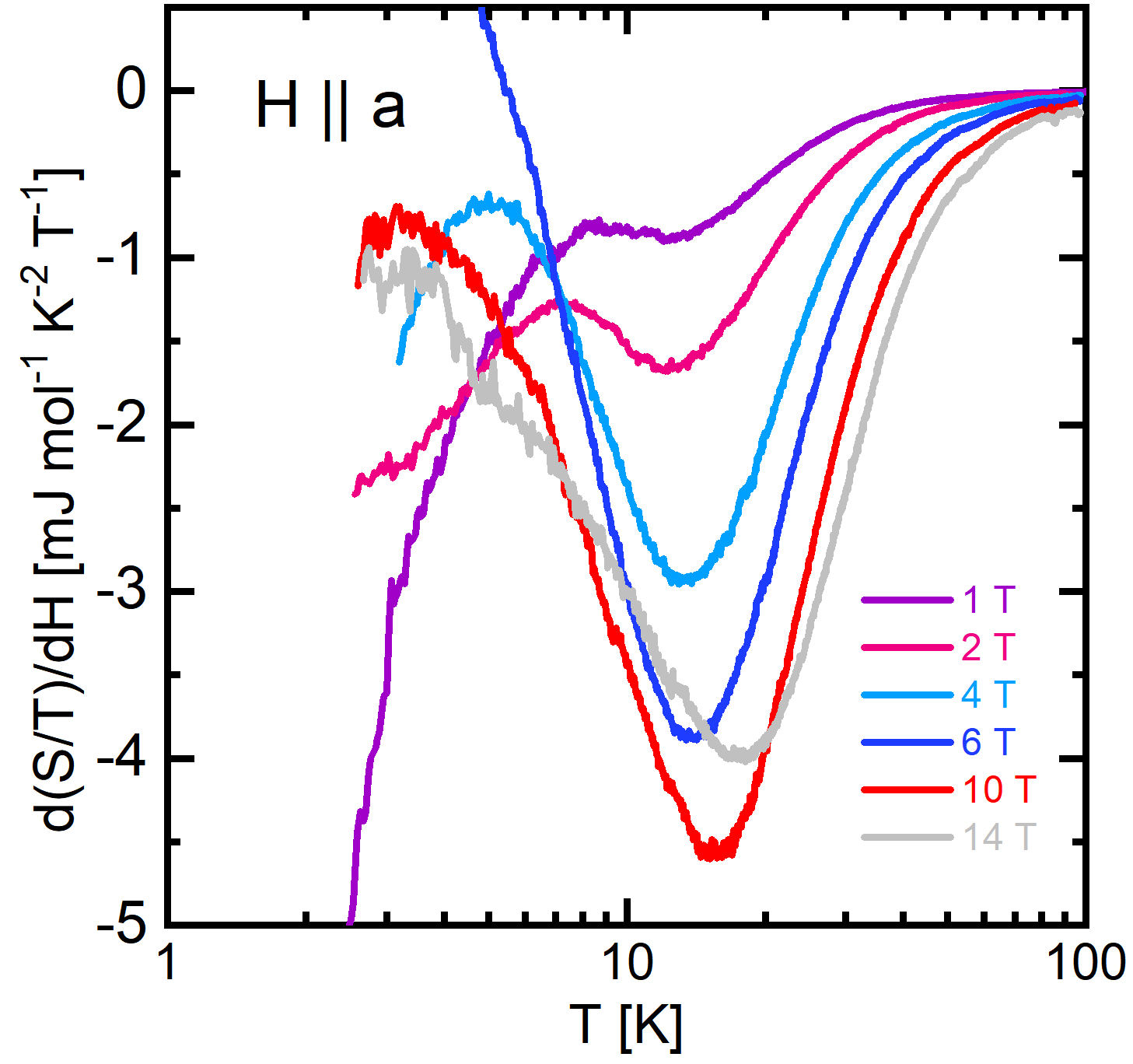}
%-----------------------------------------------------------
\caption{$\partial(S/T)/\partial H$ obtained from magnetization measurements for fields along the $a$-axis using equation \ref{eq:maxwell}}      
\end{figure}

Fig.S2 illustrates the temperature dependence of the H-derivative of the entropy inferred from our magnetization data using the Maxwell relation

\begin{equation}
\left(\frac{\partial M}{\partial T}\right) = \left(\frac{\partial S}{\partial H}\right)
\label{eq:maxwell}
\end{equation}

A minimum is observed around 12 K which increases in magnitude, broadens significantly, and shifts to higher temperature with increasing magnetic field $H||a$. This provides further thermodynamic evidence of the field-induced behavior of $T^*(H)$ observed both in thermal-expansion and resistivity measurements as explained in detail in the main text. 

\begin{figure*}[!t]
\centering
%-----------------------------------------------------------
\includegraphics[width=\textwidth]{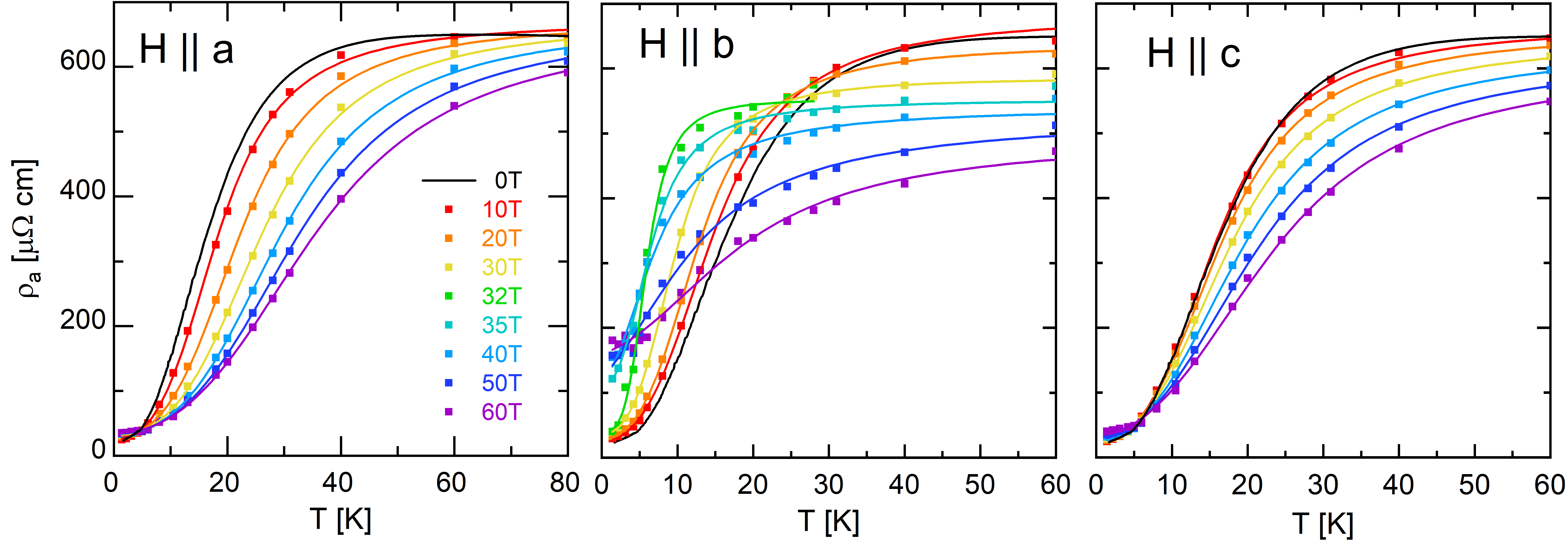}
%-----------------------------------------------------------
\caption{Resistivity data from Ref \onlinecite{Knafo19} measured along the a-axis with fields applied along all three crystallographic directions. Smooth fits are obtained with the fit function given in equation \ref{eq:fit}.}      
\end{figure*}

Figure S3 shows the temperature dependence of the $a$-axis resistivity inferred from the pulsed-field data of Knafo et al. in Ref \onlinecite{Knafo19} for fields applied along the 3 orthorhombic axes. To provide the smooth derivative $\partial(\rho)/\partial(T)$  shown in Figs 2\textbf{d-f} of the main paper, we have fitted the data to the function:

\begin{equation}
\rho_a(T,H)=\rho_0(H) + D(H)arctan\left[\left(T/T_0(H)\right)^n\right]
\label{eq:fit}
\end{equation}

where $\rho_0(H)$, $D(H)$, $T_0(H)$, and $n$ are free fit parameters (solid lines in Fig.S3). This function fits the data extremely well, but we do not attach any physical significance to this function.  It is only used to obtain a smooth derivative. The maximum in the derivative of this fit function was then used to obtain $T^*(H)$ plotted in the phase diagram of figure 2\textbf{k} in the main text.

\begin{figure*}[!h]
\centering
%-----------------------------------------------------------
\includegraphics[width=\textwidth]{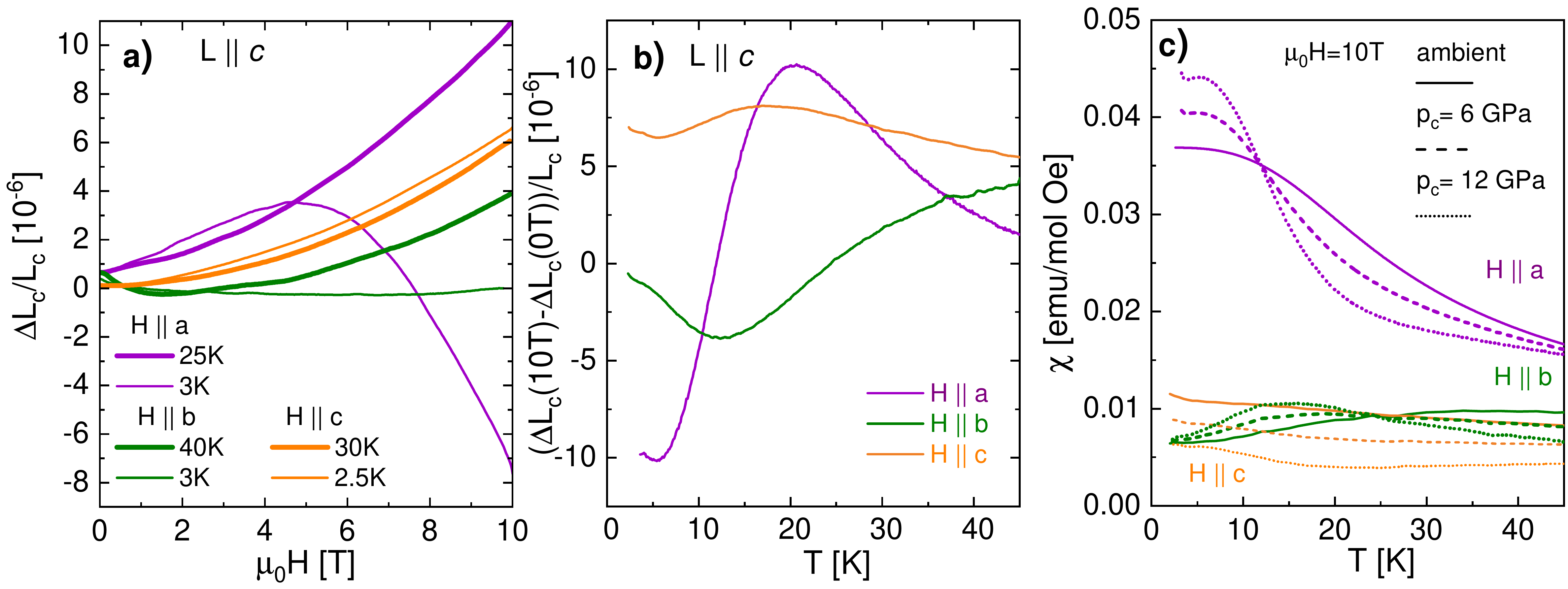}
%-----------------------------------------------------------
\caption{\textbf{a} c-axis magnetostriction for applied fields along all three crystallographic axes measured at a temperature below and above $T^*$. \textbf{b} Temperature dependent magnetostriction obtained by the difference of thermal expansion in 0T and 10T for fields applied along the $a$, $b$, and $c$-axis. \textbf{c} Predicted change of the zero-pressure susceptibility (solid lines) under the application of 6 (dashed) and 12 GPa (dotted line) $c$-axis uniaxial pressure for fields applied along the three crystallographic directions.}      
\end{figure*}

$C$-axis magnetostriction measurements were performed for fields along all three principal crystallographic directions both for high ($T>T^*$) and low ($T<T^*$) temperatures, as shown in Figure S4 \textbf{a}. We choose the c-axis because it shows the largest effect in the thermal expansion at T*. The magnitude of magnetostriction is relatively small ($\Delta L/L<10^{-5}$) compared to other heavy-fermion systems and, except for the $a$ axis fields at 3 K, varies with $H^2$ as expected for a Fermi liquid.  This is the expected magnetostriction behavior for a linear field dependence of the magnetization.
%The peak observed for low fields ($\mu_{0}|H|<1\mathrm{T}$) along the $b$ axis is caused by the background signal of the dilatometry cell.
For $a$ axis field there is a change of sign in the slope of the magnetostriction at about 5 T, which is most likely related to the Lifshitz transition observed in the thermopower \cite{Niu20}. Magnetostriction probes the change in magnetisation with pressure via the Maxwell relation  $\partial \vec{M}/\partial p = - \partial V/\partial \vec{B}$, with  $p$ the hydrostatic pressure. For uniaxial pressure, the appropriate relation assumes the tensor form
\begin{align}\label{}
   \lambda_{ij} = \frac{1}{L_i}\frac{\partial L_i}{\partial B_j} = -\frac{1}{V}\frac{\partial M_j}{\partial p_i}
\end{align}
with $i,j \in \{a,b,c\}$ the crystallographic directions, we obtain for our experimental geometry ($i=c$),
\begin{align}\label{eq:susceptibility}
   \frac{\lambda_{cj}}{B_j} = - \frac{1}{\mu_0}\frac{\partial \chi_j}{\partial p_c}
\end{align}
The left-hand side is experimentally accessible and hence allows to probe the pressure dependence of the magnetic susceptibility $\chi_{j} = V^{-1} (\partial M_{j} / \partial H_{j})$. The predicted change in susceptibility upon pressure is small, negative and constant in field for high temperatures along all three directions. At low temperatures the expected change in susceptibility would be small and positive for fields along the c-axis, zero for fields along b and changes sign for fields along the a-axis from negative to positive from below to above 5T respectively.

In order to probe the temperature dependence of the magnetostriction, we have measured the difference in thermal expansion in zero field and in 10T from 2K to 50K. This difference is shown in figure S4 \textbf{b} for field orientations along the $a$, $b$, and $c$ axes. This difference allows us to evaluate the change in magnetic susceptibility at 10 T for c-axis uniaxial pressure for the temperature range 2K  and 45K using equation \ref{eq:susceptibility}. In order illustrate our results, the magnetic susceptibility at 10 T as measured (pc=0) is plotted together with the extrapolated (from the zero-pressure slope obtained from S4 \textbf{b}) values for pc=6 GPa and 12 GPa in Fig. S4 \textbf{c}. The expected pressure effect is very distinct for the three field directions. $\chi_a(T)$ increases in the low-T Fermi-liquid regime ($T<5K$) with $c$-axis pressure, consistent with the increased Sommerfeld coefficient predicted from the thermal expansion (see main text). Further, the low-T increase of $\chi_a(T)$ sharpens significantly similar to the sharpening observed in resistivity \cite{Ran20}. For $\chi_c(T)$ a monotonic temperature-independent decrease is found, while a clear shift of $T^*$ is observed for $\chi_b(T)$. Figure S4\textbf{c} further suggests a switching of the magnetic hard direction from the $b$-axis to the $c$-axis under pressure, which has been recently observed for hydrostatic pressure \cite{Li2021}. This demonstrates that the physical changes observed under hydrostatic pressure are largely due to the c-axis pressure component.

\bibliographystyle{apsrev4-2}